\documentclass[useAMS,usenatbib]{mn2e}

\usepackage{color}
\usepackage{graphicx}
\usepackage{times}
\usepackage{amssymb}
\usepackage{amsmath}

\title[Galaxy Infall onto SDSS Groups]{The Redshift--Space Cluster--Galaxy
    Cross--Correlation Function: I. Modeling Galaxy Infall onto Millennium
    Simulation Clusters and SDSS Groups}
\author[Zu et al.]{ 
\parbox{\textwidth}{
Ying Zu$^1$\thanks{E-mail: yingzu@astronomy.ohio-state.edu},
David H. Weinberg$^{1}$
}
\vspace*{4pt} \\
${1}$~Department of Astronomy and CCAPP, The Ohio State University, 140 W. 18th Avenue,
Columbus, OH 43210, USA\\
}

\def\xiggs{\xi^{s}_{gg}}
\def\xicgs{\xi^{s}_{cg}}
\def\xicgr{\xi^{r}_{cg}}
\def\hmpc{h^{-1}\mathrm{Mpc}}
\def\hkpc{h^{-1}\mathrm{kpc}}
\def\hmsol{h^{-1}M_\odot}
\def\fvlos{f(v_{\mathrm{los}}|r_p, y)}
\def\p2d{P(v_r, v_t)}
\def\mvir{\mathcal{G}}
\def\minf{\mathcal{T}}
\def\rsh{r_{\mathrm{sh}}}
\def\rmi{r_{\mathrm{mi}}}

\def\vrc{v_{r, c}}
\def\f{f_{\mathrm{vir}}}
\def\sv{\sigma_{\mathrm{vir}}}
\def\sr{\sigma_{\mathrm{rad}}}
\def\st{\sigma_{\mathrm{tan}}}
\def\s0{\sigma_{0}}
\def\rpic{r_{\pi, c}}
\def\dof{\mathtt{dof}}

\begin{document} 
\date{\today} \maketitle
%
\begin{abstract}
The large scale infall of galaxies around massive clusters provides a
potentially powerful diagnostic of structure growth, dark energy, and
cosmological deviations from General Relativity. We develop and test
a method to recover galaxy infall kinematics~(GIK) from measurements
of the redshift--space cluster--galaxy cross--correlation function
$\xicgs(r_p,r_\pi)$.  Using galaxy and halo samples from the Millennium
simulation, we calibrate an analytic model of the galaxy kinematic
profiles comprised of a virialized component with an isotropic Gaussian
velocity distribution and an infall component described by a skewed 2D
$t$-distribution with a characteristic infall velocity $v_{r,c}$ and separate
radial and tangential dispersions.  We show that convolving the real-space
cross-correlation function with this velocity distribution accurately
predicts the redshift-space $\xicgs$, and we show that measurements of
$\xicgs$ can be inverted to recover the four distinct elements of the GIK
profiles. These in turn provide diagnostics of cluster mass profiles, and
we expect the characteristic infall velocity $v_{r,c}(r)$ in particular
to be insensitive to galaxy formation physics that can affect velocity
dispersions within halos. As a proof of concept we measure $\xicgs$ for
rich galaxy groups in the Sloan Digital Sky Survey and recover GIK profiles
for groups in two bins of central galaxy stellar mass. The higher mass bin
has a $v_{r,c}(r)$ curve very similar to that of $10^{14}\,\hmsol$ halos in
the Millennium simulation, and the recovered kinematics follow the expected
trends with mass. GIK modeling of cluster--galaxy cross--correlations can be
a valuable complement to stacked weak lensing analyses, allowing novel tests
of modified gravity theories that seek to explain cosmic acceleration.
\end{abstract}
\section{Introduction}
\label{sec:intro}

As the largest bound systems in the universe, galaxy clusters carry imprints
of cosmic growth via the distribution and motion of their surrounding dark
matter and galaxies~\citep[see][and references within]{allen2011,kravtsov2012}.
They can therefore play a powerful role in testing theories for the origin
of cosmic acceleration, complementing geometrical probes such as supernovae
and baryon acoustic oscillations. The key uncertainty in this approach is
calibration of the relation between cluster observables~(e.g., X-ray luminosity
or temperature, optical galaxy richness, Sunyaev--Zel'dovich decrement)
and halo mass. Stacked weak lensing has emerged as a robust approach
to this problem because it is unaffected by the baryonic physics of the
intracluster gas~(\citeauthor{mandelbaum2006} 2006; \citeauthor{sheldon2009}
2009; \citeauthor{oguri2012} 2012, see \citeauthor{weinberg2012} 2012 for a
review of this approach in the broader context of dark energy studies). Galaxy
infall patterns~\citep{gunn1972, ryden1987, regos1989} offer an alternative
probe of cluster mass profiles, which may prove a valuable complement
to weak lensing if it can be implemented in a way that is insensitive to
uncertainties of galaxy formation physics. The redshift--space cluster--galaxy
cross--correlation function, $\xicgs$, is a comprehensive characterization of
the statistical relation between clusters and galaxies, influenced by both
the real--space cross--correlation and the peculiar velocities induced by
the cluster gravitational potential. This paper is the first of a three-part
series which will describe the modeling of $\xicgs$ and investigate its
diagnostic power for cluster mass calibration and constraining cosmology.

In cluster--centric coordinates, the average galaxy kinematics are
the result of competition between the cluster potential and Hubble
expansion. At small distances, virial motion dominates, making the galaxy
distribution elongated along the line-of-sight~(LOS) direction~(a.k.a. the
``Fingers-of-God'' effect; FOG)~\citep{jackson1972}. At larger distances, the galaxy kinematics
become dominated by radial infall, with a typical turn--around radius of
several $\hmpc$,\footnote{Here $h\equiv H_0/100\,\mathrm{kms}^{-1}\mathrm{Mpc}^{-1}$}
where the characteristic infall velocity is equal to the Hubble flow.
This coherent infall produces a squashing distortion in $\xicgs$ at large
scales, often referred to as the Kaiser effect~(\citeauthor{kaiser1987} 1987;
see also~\citeauthor{sargent1977} 1977). These two redshift space distortion~(RSD)
effects~(see \citeauthor{hamilton1998} 1998 for a pedagogical review) are also
seen in the redshift--space galaxy auto--correlation functions~\citep[e.g.,][and
references within]{reid2012}, but since galaxies feel a much stronger central
potential near clusters than in the field, both small scale dispersion and
large scale infall are strongly enhanced in the $\xicgs$ case.

The strong RSD in $\xicgs$ allows reconstruction of the average galaxy
kinematics around clusters, which are in turn
 determined by the average cluster mass profiles. For clusters of fixed mass
 $M$, $\xicgs(r_p, r_\pi)$
at projected separation $r_p$ and LOS separation $r_\pi$ can be derived by
convolving the real--space cross--correlation function $\xicgr$ with the LOS
velocity distribution function $\fvlos$~\citep{peebles1980, fisher1995},
\begin{equation}
\xicgs(r_p, r_\pi) +1 = \left[\xicgr\left(\sqrt{r_p^2+y^2}\right) + 1\right] \ast \fvlos,
\label{eqn:convolution}
\end{equation}
where $y$ is the LOS separation in real space.  The real--space $\xicgr$ has
a roughly power--law form, so the strongest features in $\xicgs$ arise
largely from $\fvlos$. In this paper, we first develop an analytic
description of the average galaxy infall kinematics~(GIK) around cluster--mass
halos found in the Millennium simulation~\citep{springel2005}. We show that
applying this model to Equation~\ref{eqn:convolution} accurately reproduces
the simulated $\xicgs$, and we show that the analysis can be reversed to
infer the correct GIK from measurements of $\xicgs(r_p, r_\pi)$. As an
illustrative application to observed data, we measure $\xicgs$ for galaxy
groups~\citep{yang2007} identified in the Sloan Digital Sky Survey~(SDSS;
\citeauthor{york2000} 2000) and apply our model to infer the infall kinematics.

At small scales, the redshift--space distribution of $\xicgs$ depends mainly
on the velocity dispersion profile of the virialized cluster component. At
large separations, one must consider the mean radial velocity profile and
the profiles of radial and tangential velocity dispersions. (In fact, we
will show that the velocity distributions are significantly non--Gaussian
and exhibit internal correlations, all of which must be modeled to describe
$\xicgs$ accurately.)  All four of these profiles can be reconstructed from our
$\xicgs$ modeling techniques, and all four of them provide diagnostics of cluster
mass. We are particularly hopeful that the mean radial flow will prove to
be a tool for inferring average mass profiles that is insensitive to galaxy
formation physics. The cluster velocity dispersion profile can be affected at
the $10\--20\%$ level by orbital anisotropy and potentially by other effects
such as preferential destruction of galaxies that pass near the cluster center
or incomplete relaxation of substructures. The velocity dispersions of galaxies
within infalling halos may also be affected by biases arising from galaxy
formation physics. However, the mean velocity of galaxies in an infalling halo
should, on average, match the mean velocity of the halo itself, as galaxies and
dark matter are affected by the same gravitational potential. We will test this
conjecture in future work.

\begin{figure*}
\centering
\resizebox{1.0\textwidth}{!}
{\includegraphics{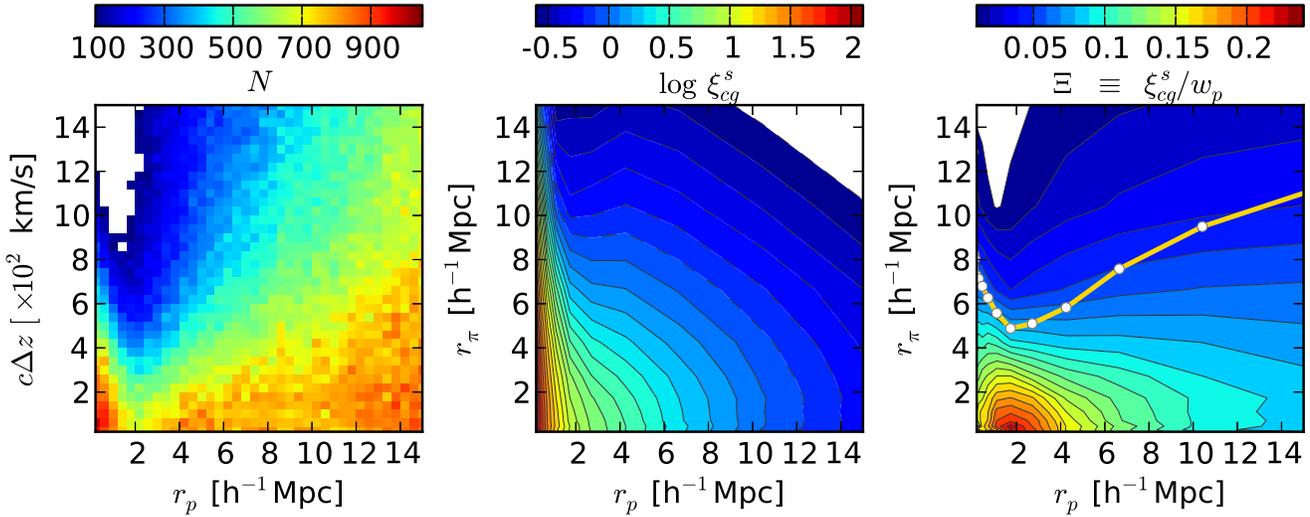}}
\caption{Three representations of the
average galaxy distribution around Millennium clusters with mass between
$1.259-1.585\times10^{14}M_\odot$ in the redshift--projected separation
plane. {\it Left panel}: Number of stacked galaxies in $r_p$-$c\Delta z$
cells. {\it Middle panel}: Cluster--galaxy correlation function in redshift space
$\xicgs$. {\it Right panel}: $\xicgs$ normalized by the projected
cluster--galaxy correlation function $w_p$ at each $r_p$. The yellow solid U-shape
curve delineates the characteristic scale of $r_\pi$ at which $\xicgs$
drops by one e--fold at fixed $r_p$, relative to its value at $r_\pi=0$. The contour levels for each panel are
colour--coded by the top colour bar.}
\label{fig:xirsdemo}
\end{figure*}

Around individual clusters, the galaxy distribution in the ($r_p$,
$r_\pi$) plane shows a distinctive trumpet--shaped pattern, and the
distribution in $r_\pi$ at fixed $r_p$ often shows a caustic--like
discontinuity~\citep{diaferio1997}. N-body simulations suggest that the caustic
location provides a direct measure of the escape velocity profile, which
can in turn be converted to a cluster mass profile~(\citeauthor{diaferio1999}
1999; see \citeauthor[][]{serra2011} 2011 for the current state of the art).
\cite{rines2003} have used this technique to infer cluster mass profiles
extending beyond the virial radius~\citep[also see][]{rines2006, geller2012, rines2012}. However, measurements for any individual cluster are affected by
galaxy shot noise and by departures from spherical symmetry~\citep{white2010},
and it is not clear whether averaging measurements from multiple clusters will
yield an unbiased mean result. Our initial motivation for this study was, in
part, to generalize the ``caustic method'' to the case where an overlapping
cluster survey and galaxy redshift survey provide a large total number of
cluster--galaxy pairs, even though the numbers around an individual cluster
may be too small for caustic detection. Although we will draw connections
between $\xicgs$ and the trumpet--shaped patterns of the caustic method,
our approach ultimately does not rely on finding discontinuities in the
data or identifying them with the escape velocity profile. Instead it
relies on the general predictions of velocity distributions around massive
halos in cosmological N-body simulations. Previous analyses of $\xicgs$
include the study of~\cite{croft1999} using APM galaxy clusters and the
study of~\cite{li2012} using velocity dispersion profiles of groups in the
SDSS. Both studies assume Gaussian LOS velocity distributions with a mean
radial infall profile. Here we adopt a more comprehensive approach to model
not only the first two moments of velocity distributions, but the full GIK
from the inner $1\,\hmpc$ to scales beyond $40\,\hmpc$.

One can imagine two somewhat different ways to go from GIK modeling of
$\xicgs$ to constraints on cosmological parameters. One is to calibrate the
average masses of clusters in bins of cluster observables, then combine this
calibration with cluster counts to extract cosmological constraints via the
halo mass function, analogous to approach of \cite{rozo2010} using stacked
weak lensing.  The second is to extract constraints directly from $\xicgs$
itself, marginalizing over uncertainties in galaxy bias. Our primary goals
in this paper are to understand the physical origin of the features in
$\xicgs$ and develop a method for inferring GIK statistics from $\xicgs$
measurements. In the second paper of this series we will investigate GIK as a
method for measuring mean cluster mass profiles, with particular attention to
galaxy bias effects. In the third paper we will investigate the cosmological
information that can be derived from $\xicgs$, in comparison to and combination
with cluster weak lensing~\citep{rozo2010, zu2012}.

In the context of standard dark energy models, $\xicgs$ and stacked weak
lensing analyses involve different types of systematic uncertainties. In
the context of modified gravity theories of cosmic acceleration, they
also provide distinct information. Gravitational lensing and the motions
of non--relativistic tracers are affected by different combinations
of gravitational potentials, which are equal in GR but unequal in many
modified gravity theories~\citep{jain2010}.  Furthermore, transitions
between ``unshielded'' and ``shielded'' regimes of modified gravity can
produce distinctive features in the density and velocity fields around
clusters~\citep{lombriser2012, lam2012}, which may reveal themselves as
unusual features in $\xicgs$. We will investigate this sensitivity to
alternative gravity theories in future work.

We begin our study by characterizing the GIK mass halos in the Millennium
simulation with a compact analytic description. In \S\ref{sec:xicgs}
we show that the GIK model, in combination with the real--space $\xicgr$,
accurately reproduces the Millennium $\xicgs$. In \S\ref{sec:mcmcmill} we show
that the GIK parameters can be reconstructed from measurements of $\xicgs$, and
in \S\ref{sec:mcmcsdss} we apply our methodology to measurements of $\xicgs$ for
SDSS groups. We summarize our results and discuss prospects for this approach in
\S\ref{sec:con}.

\section{$\mathbf{\xicgs}$ and Galaxy Infall Kinematics Around
Millennium Simulation Halos}
\label{sec:gik}

To investigate the average galaxy velocity distribution as a function of
cluster--centric radius, we make use of the semi--analytic model~(SAM) galaxies
inside the Millennium simulation~\citep{springel2005}, which evolves $2160^3$
dark matter particles with $M_p=8.6\times10^8\,\hmsol$ in a periodic box
500$\hmpc$ on a side.\footnote{All the distances in the paper are in comoving
units, relative to halo centers. All the kinematics are relative to halo
center-of-mass velocities.} The underlying cosmological model is inflationary
cold dark matter with a cosmological constant~($\Lambda$CDM), though the
values of the matter density $\Omega_m$ and power spectrum normalization
$\sigma_8$ are somewhat high compared to current estimates.  The SAM galaxies
are then constructed by assigning an empirical galaxy formation recipe along
the merger trees of dark matter halos~\citep{de_lucia2007}. Since the galaxy
kinematics are primarily determined by the long--range gravitational forces that
are unaware of the detailed baryonic physics inside galaxies, we can mostly
treat the kinematics of galaxies and their host sub--halos as equal. The
only exception is within the cluster virial radius, where the prescriptions
for dynamical friction may alter the kinematics of galaxies after their host
sub--halos fail to survive above the resolution limit of simulation because of
tidal truncation~\citep[][]{ghigna2000,gao2004,kravtsov2004}. The particular
dynamical friction prescription adopted in~\cite{de_lucia2007} uses a variant
of the classic formula from~\cite{binney1987} to calculate the friction time
scale. After this time scale, galaxies are assumed to merge onto the central
galaxies of the main halo. The uncertainties in the dynamical friction time
scale are significant, but they only affect the very inner region that our
analysis is insensitive to. Therefore, we believe the galaxy kinematics
measured from the SAM sample are representative of $\Lambda$CDM+General
Relativity~(GR) models. We will comment on the potential impacts of different
dynamical friction prescriptions on our GIK model in Paper II.

To obtain a decent signal-to-noise ratio, we select SAM galaxies with absolute
SDSS $r$-band magnitude $M_r < -18$ at $z=0.1$, which produces a sample size
of $6791721$ galaxies. We also repeated our kinematics measurements using a
brighter sample with $M_r<-20$ and found negligible differences except for
relatively larger noise. We nonetheless will investigate the dependence of GIK
on galaxy type and luminosity in Paper II. We selected clusters\footnote{We
use the terms clusters, groups, and dark matter halos nearly interchangeably
throughout the paper, though ``groups'' should be understood to refer to the
low end of the cluster mass range.} in six mass bins with $0.1$ dex in bin width, which corresponds to at most $25\%$ difference in mass, and $\leq 8\%$ 
difference in overall velocity amplitude within each bin~(assuming
characteristic velocities $v\propto M^{1/3}$). The narrow bin width ensures
that the scatter we measure in the galaxy velocities within each mass bin
is intrinsic, instead of extrinsic scatter induced by the scatter in cluster
masses within that bin. The mass bins are indicated by the top left panel of
Fig.~\ref{fig:parfit}~(discussed further below). The clusters are identified
by spherical overdensity in the simulation, and the cluster mass $M_c$
is defined as the mass inside a sphere with enclosed density $200$ times
the mean matter density $\Omega_m$. Throughout our analysis, we choose our
fiducial mass bin to be $M_c = 1.259\--1.585 \times 10^{14}\,\hmsol$.

\begin{figure}
\centering
\resizebox{0.45\textwidth}{!}
{\includegraphics{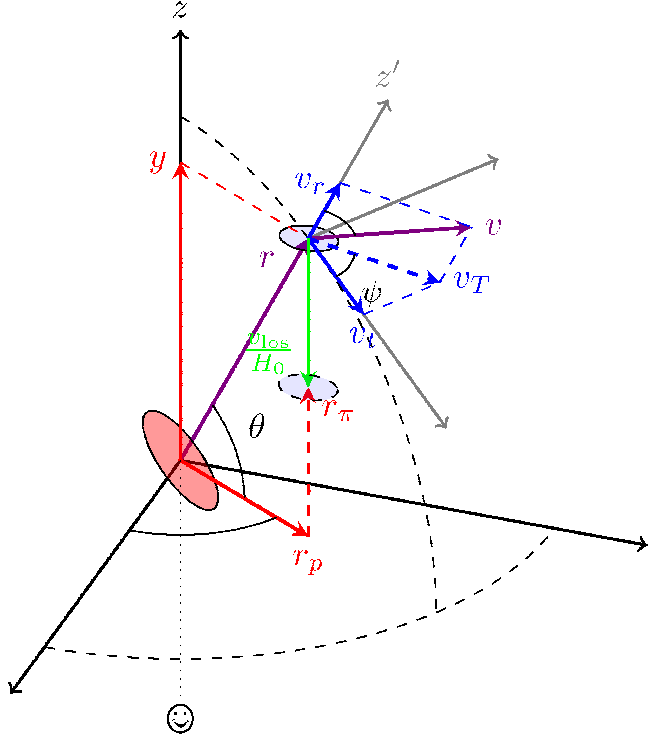}}
\caption{
    Definition of the various position and velocity vectors in the
    cluster--centric coordinate~(black solid axes), where red and blue
    solid ellipses represent the central and satellite galaxies. The LOS
    direction is along the z-axis. Gray solid axes indicate the local
    cartesian coordinate carried by the satellite. The blue dashed ellipse
    indicates the position of satellite in redshift space after displacement
    caused by $v_\mathrm{los}$.
}
\label{fig:coord}
\end{figure}

Fig.~\ref{fig:xirsdemo} compares three different ways of illustrating the
average distribution of galaxies around clusters in redshift space, using
our fiducial cluster mass bin as an example. The left panel simply shows the
stacked number counts in cells of equal area on the redshift--$r_p$ plane.
As mentioned in the introduction, this resembles the traditional way in which
the ``caustic'' curve is identified for individual clusters --- we can see an
enhancement of the galaxy number distribution at small redshift separations
forming the trumpet-shaped pattern. At small $r_p$, the galaxy distribution
is stretched along the LOS by virial motions, but at $r_p\sim2\,\hmpc$ it is
highly compressed along the LOS by infall. While there are strong gradients in
$r_\pi$ at each $r_p$, there is not an obvious line of discontinuity, perhaps
because caustics of individual clusters are washed out by scatter in the stack.

Note that although the cells in this representation have equal area, those at
large $r_p$ have larger volumes in 3-dimensional redshift space because each
cell represents a cylindrical ring with radius $r_p$~($V_{\mathrm{cell}}\propto
r_p^2$). The middle panel shows the central quantity in this paper, the
redshift--space cluster--galaxy correlation function $\xicgs$. Since it
is equivalent to the overdensity of galaxies around clusters, the dominant
feature we see, other than the FOG and Kaiser effects, is the declining trend
of overdensity with distance. To highlight the cluster RSD effects at fixed
$r_p$, in the right panel we plot the contours of $\Xi$, which is defined
as the ratio between $\xicgs(r_p, r_\pi)$ and the projected correlation
function $w_p(r_p)$ at given $r_p$. Since $w_p(r_p)$ is the integral of
$\xicgs$ along $r_\pi$ at fixed $r_p$,\footnote{$w_p(r_p) =
    \int_{-\infty}^{+\infty}\xicgr\left(\sqrt{r_p^2+y^2}\right)d
y$. In practice, we cut off the integration at $y=\pm 40\,\hmpc$.}
the ratio $\Xi$ has no information that is not already in $\xicgs$, but by
scaling out the radial trend it highlights the RSD effects.
The highest peak for $\Xi$ is no longer at the origin~(i.e., cluster
center), but migrates horizontally to the region around $r_p\simeq2\hmpc$,
meaning that the relative distribution of galaxies along $r_\pi$ is the
most compact at $r_p\simeq2\,\hmpc$ and becomes more spread at both small
and large scales. The yellow solid curve in the right panel quantifies this
compactness --- it shows the characteristic LOS distance $r_{\pi,c}$ at which
the amplitude of $\Xi$ drops by $1/e$ from $\Xi(r_\pi=0)$ at each $r_p$. The
curve has a characteristic U-shape, indicating a more compact distribution of
galaxies at intermediate projected radii. Note that because $\Xi\propto\xicgs$
at each $r_p$, this U-shaped curve would occupy the same locus in the $\xicgs$
plot of the middle panel, although we do not show it there.

\begin{figure*}
\centering
\resizebox{1.0\textwidth}{!}
{\includegraphics{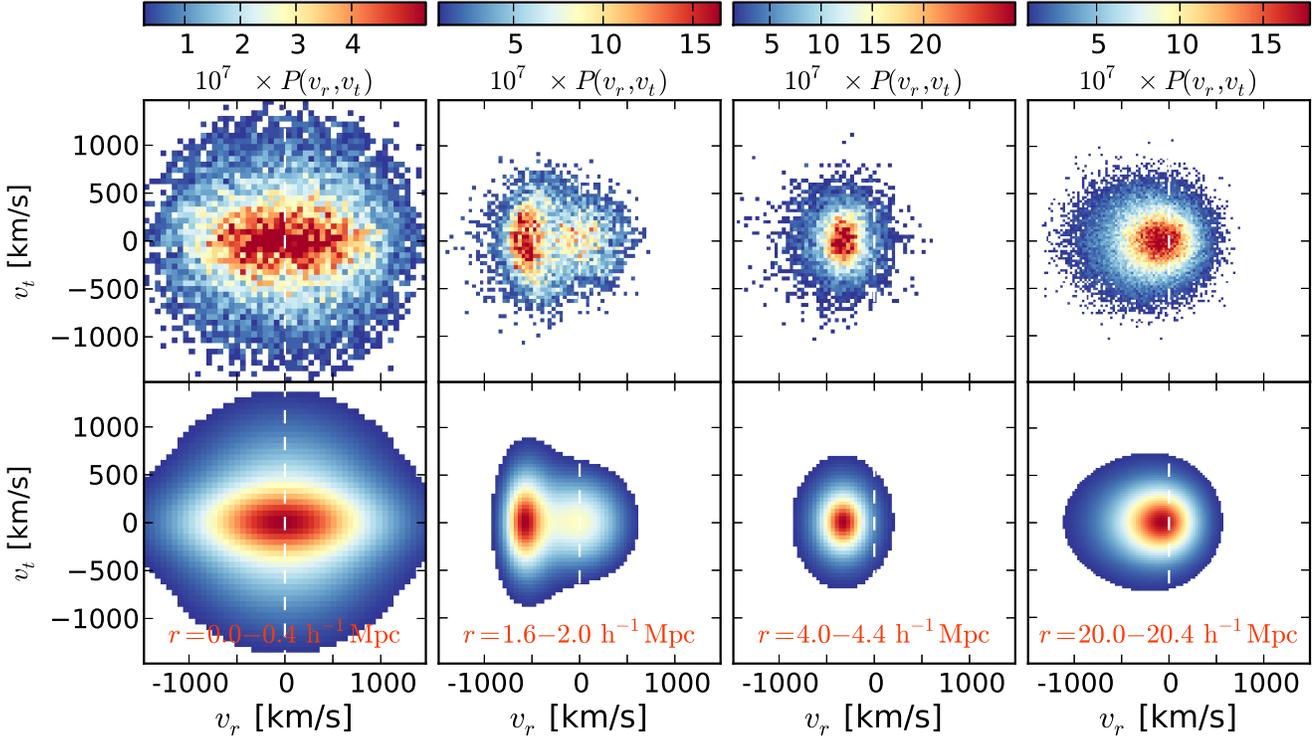}}
\caption{Joint probability distributions of radial and
tangential velocities $P(v_r, v_t)$ from the
simulation~(top panels) and
the best--fit using our GIK model~(bottom panels), in four different
radial bins marked at the bottom of each panel. The colour scales used by
panels in the same column are identical, indicated by the colour bar on top.
}
\label{fig:v2dfit}
\end{figure*}

What determines the location of this characteristic U-shaped curve?  It bears
some resemblance to the caustic curve for individual clusters, where the
fall--off of the galaxy distribution on the redshift--$r_p$ diagram is
believed to be the natural boundary imposed by the escape velocities at
different radii.  Indeed, in Fig.~\ref{fig:xirsdemo} the U-shaped curve
in the right panel is similar in shape to isodensity contours in the left
panel, with the normalization by the mean expected galaxy number removing
the somewhat arbitrary geometric weighting~($\propto r_p^2$) of the straight
number--count diagram.  However, with no clear discontinuity in $\xicgs$
or $\Xi$, we cannot relate this curve directly to escape velocities. To
explain it quantitatively, we need to understand the average galaxy velocity
distribution function around clusters, and in particular, $\fvlos$.

\begin{figure*}
\centering
\resizebox{1.0\textwidth}{!}
{\includegraphics{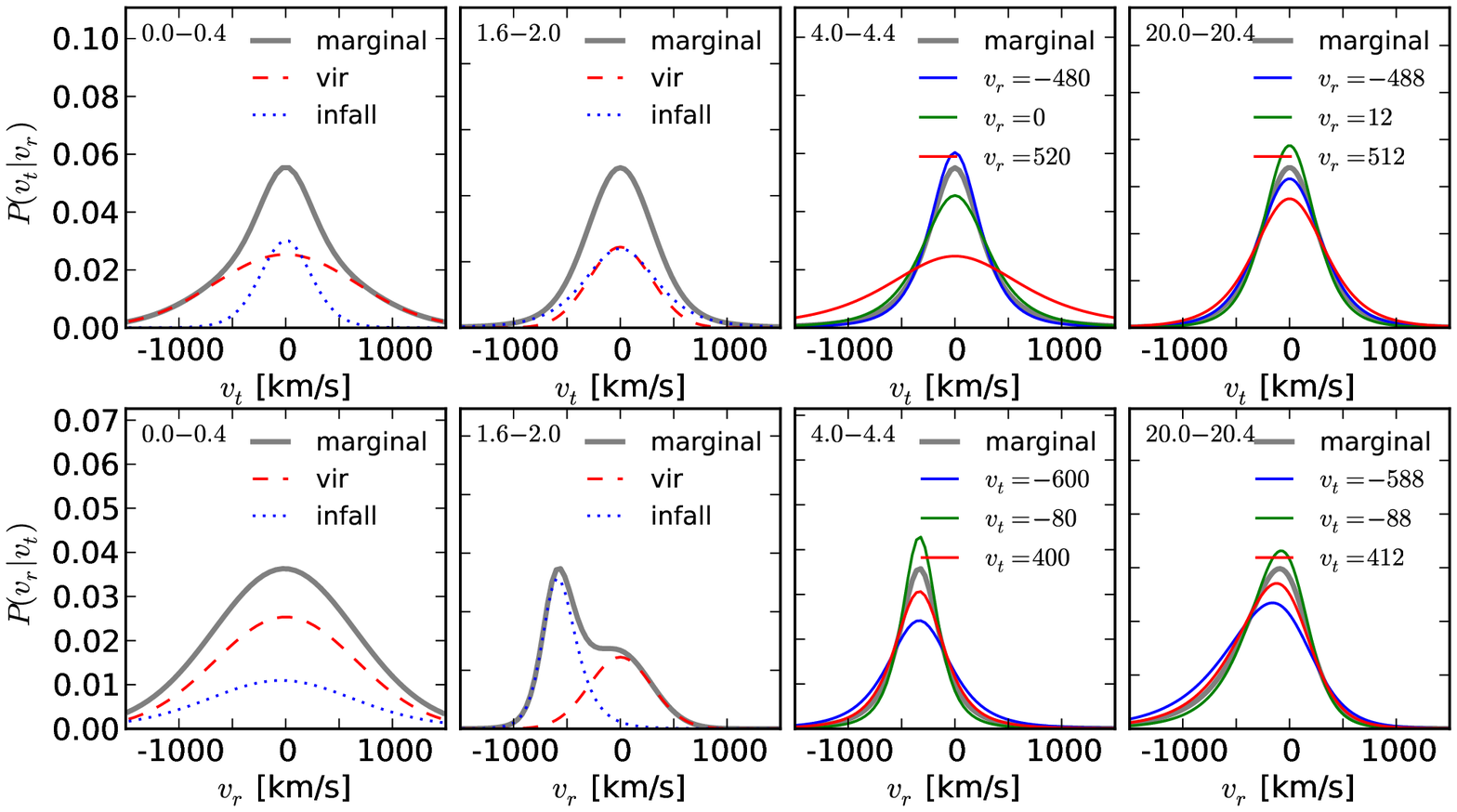}}
\caption{Probability distributions of tangential~(top panels) and
radial velocities~(bottom panels) at the same four radial bins as in
Fig.~\ref{fig:v2dfit}. In each panel, the gray solid curve shows the 1D marginal
distribution. For the left two panels, red dashed and blue dotted curves
show the relative contribution from the virialized and infall components,
respectively; for the right two top~(bottom) panels, blue, green, and red
curves show the conditional probability distributions of $v_t$~($v_r$)
at three different $v_r$~($v_t$), as labeled.}
\label{fig:pvcond}
\end{figure*}

For each mass bin of clusters, we stacked all the galaxies in the
cluster--centric coordinate to produce a synthetic cluster of that mass bin.
Although individual clusters vary in shape and lumpiness, the synthetic
clusters are isotropic and smooth by construction. Therefore, in order to fully
describe $\fvlos$, we only need to measure the joint probability distribution
function~(PDF) of radial velocity $v_r$ and ``half'' the tangential velocity
component $v_{t}$, at each radius $r$, $P(v_r, v_t |r)$, so that
\begin{equation}
\fvlos  = \int_{-\infty}^{+\infty} P\left(v_r, v_t=\frac{v_\mathrm{los} - v_r
\sin\theta}{\cos\theta} \bigg| r \right) \frac{d v_r}{\cos\theta},
\label{eqn:fvlos}
\end{equation}
where $r=\sqrt{r_p^2 + y^2}$ and $\theta = \tan^{-1} y/r_p$. Here $v_t$
represents the tangential velocity~($v_T$) component that is projected in
the plane of LOS axis and galaxy position vector in the cluster--centric
frame~(see the 3D diagram in Fig.~\ref{fig:coord}). Given an isotropic
cluster, this projected component is $v_t = |v_T|\cos\psi$ where $\psi$
is randomly distributed between $-90$ and $90$ degrees, hence ``half''
of $v_T$.  To avoid redundancy, hereafter we refer to $v_t$ simply as the
``tangential velocity''.  Note that we subtract Hubble flow when defining
$v_r$, and the probability distribution of $v_t$ is symmetric about zero.

The top panels of Fig.~\ref{fig:v2dfit} shows the measured $\p2d$ for
four radial bins that represent four distinctive regimes of GIK around
clusters. Starting from the innermost radial bin~($r=0.0\--0.4\hmpc$),
the joint velocity distribution appears to be a single component ellipse
with the major axis in the radial velocity direction, indicating a
preference for radially oriented orbits that may arise from galaxies
on first infall or second passage, whose velocity directions have
not been randomized~\citep{bertschinger1985}. Going slightly further
out~($r=1.6\--2.0\hmpc$), the joint distribution is clearly resolved into
two parts, one virialized Gaussian component with zero means of $v_r$
and $v_t$, and one radial velocity--dominated component skewed toward negative
radial velocities~(infall). At $r=4.0\--4.4\hmpc$, near the turn--around
radius where infall velocity is comparable to Hubble flow, the virialized
component disappears while the infall component has a mean $v_r\simeq
400\,\mathrm{kms}^{-1}$ but no skewness. On very large scales ($r=20.0\--20.4\hmpc$),
the joint distribution is still largely infall, but skewed toward positive
velocities. The joint velocity distribution measured for any of the other radial
bins is qualitatively similar to one of the four bins shown here, or some
interpolation between two of them. To ensure an accurate description of
$\fvlos$ at $(r_p, r_\pi) <30\,\hmpc$, where the measurements of $\xicgs$
are most precise, we need to model $\p2d$ across all scales from the
inner $1\,\hmpc$ to beyond $40\,\hmpc$.

\begin{figure*}
\centering
\resizebox{1.0\textwidth}{!}
{\includegraphics{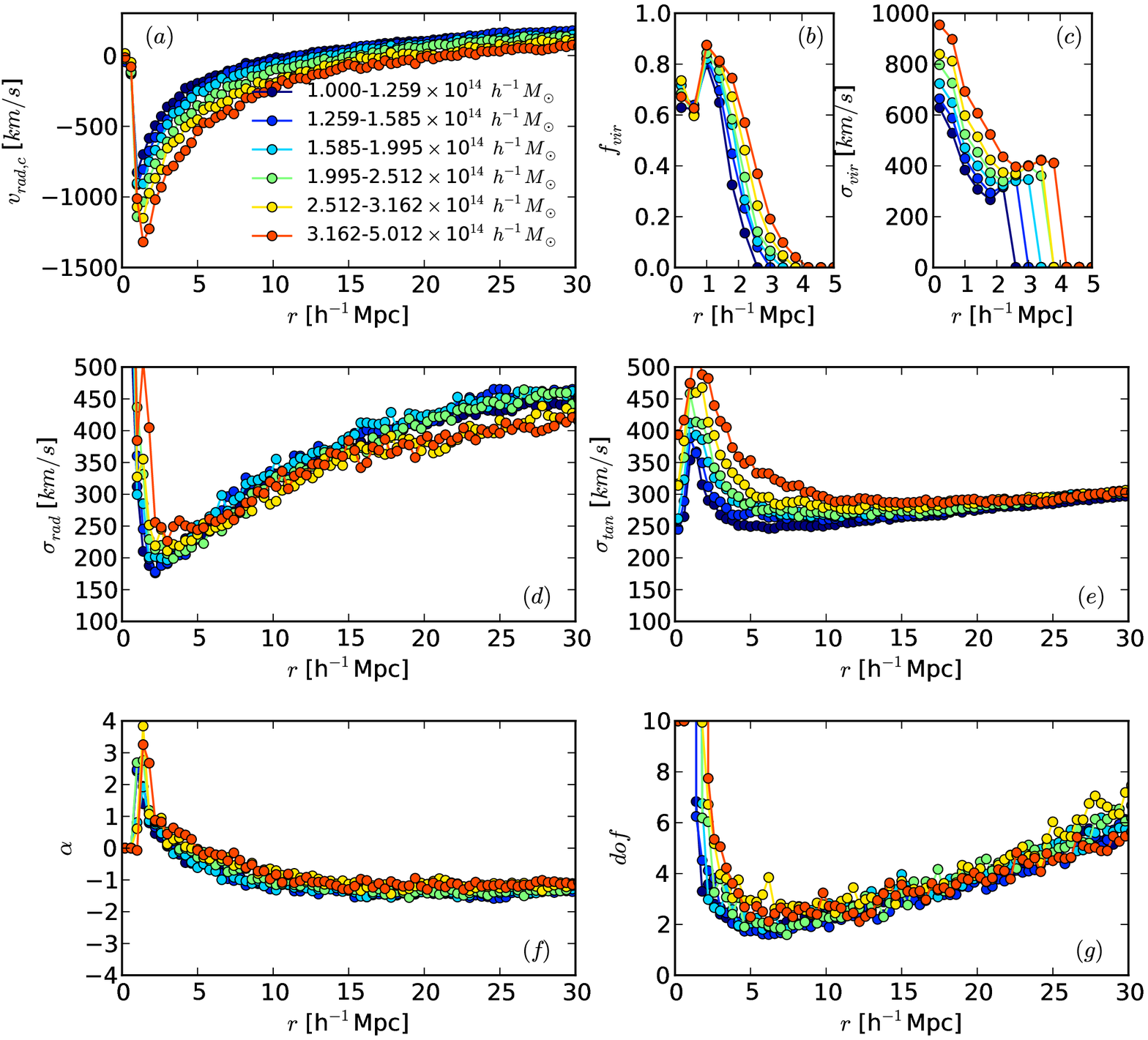}}
\caption{Best--fit GIK~(galaxy infall kinematics) model parameters as functions
    of radius for six mass bins. (a): characteristic infall velocity; (b):
    fraction of the virialized component; (c): velocity dispersion of the
    virialized component; (d): radial velocity dispersion of the infall
    component; (e): tangential velocity dispersion of the infall component;
    (f): parameter for describing skewness in the radial velocity axis; (g):
    degrees of freedom for overall kurtosis.}
\label{fig:parfit}
\end{figure*}

Motivated by the top panels of Fig.~\ref{fig:v2dfit}, we adopt a two--component
mixture model for the velocity distribution at any given cluster--centric radius
$r$, with the virialized component described by a 2D Gaussian
$\mvir$ and the infall component by a 2D skewed $t$-distribution $\minf$:
\begin{equation}
\p2d \equiv P(\mathbf{v}) = \f\cdot\mvir(\mathbf{v}) +
(1-\f)\cdot\minf(\mathbf{v}),
\label{eqn:p2d}
\end{equation}
where $\f\geq0$ is the fraction of galaxies in the virialized component, approaching zero
at large $r$. We refer to the radius beyond which $\f=0$ as the ``shock
radius'' $\rsh$, since it marks~(at least within the model) the boundary between
single--component and two--component flow. 
By definition $\mvir$ has zero mean in both radial and tangential axes, and we
find it adequate to assume equal dispersions, making $\mvir$ 
 a function of only one parameter, the virial dispersion $\sv$~(which is still
 allowed to vary with $r$). For
the infall component, describing the varying degrees of skewness and
kurtosis at different $r$ requires a functional form $\minf$ with greater complexity. We
adopt the skewed $t$--distribution parameterization from~\cite{azzalini2003},
with two parameters describing the higher order moments of the velocity
distribution~($\alpha$ and $\dof$) in addition to three parameters for the
mean and dispersions~($\vrc$, $\sr$, and $\st$). The full expression is
\begin{eqnarray}
\minf(\mathbf{v}) &=& 2 \, t_2(\mathbf{v}; \dof) \times \nonumber\\
&&
T_1\left\{\boldsymbol{\alpha}^T\boldsymbol{\omega}^{-1}(\mathbf{v}-\mathbf{\bar{v}})\cdot\left(\frac{\dof+2}{Q_v+\dof}\right);
    \dof+2\right\},
\label{eqn:skewt}
\end{eqnarray}
where $\mathbf{\bar{v}} = \bigl(\vrc, 0\bigl)$, $\boldsymbol{\alpha} =
\bigl(\alpha, 0\bigl)$, and $\boldsymbol{\omega} = \bigl(\sr, \st\bigl)$
are 2-element vectors, and $Q_v = (\mathbf{v}-\mathbf{\bar{v}})^T \Sigma^{-1}
(\mathbf{v}-\mathbf{\bar{v}})$ is a scalar where 
\begin{equation}
\Sigma = \begin{pmatrix} \sr^2&0\\ 0&\st^2 \end{pmatrix}.
\end{equation}
For the two rhs terms in Equation~\ref{eqn:skewt}, $t_2$ is the density
function of 2D $t$-variate with $\dof$ degrees of freedom,
\begin{equation}
t_2(\mathbf{v} ; \dof) = \frac{\Gamma\{(\dof+2)/2\}}{|\Sigma|^{1/2}(\pi
\,\dof)^{1/2}\Gamma(\dof/2)}\left(1 + \frac{Q_v}{\dof}\right)^{(\dof+2)/2},
\end{equation}
and $T_1(x; \dof+2)$ denotes the scalar $t$-distribution function with $\dof+2$
degrees of freedom. Generally speaking, $\boldsymbol{\alpha}$ controls the
skewness of $\p2d$ in the radial velocity direction, while $\dof$ adjusts
the kurtosis in both directions, with lower $\dof$ corresponding to longer
non--Gaussian tails. Since $\p2d$ is symmetric in the tangential velocity
axis, $\boldsymbol{\alpha}$ is reduced to one parameter $\alpha$. $\sr$
and $\st$ describe the dispersion in each direction, and $\vrc$ is the
characteristic radial velocity. Therefore, we have seven parameters in total
for $\p2d$ at every $r$: virialized fraction $\f$, velocity dispersion
of the virialized component $\sv$, characteristic infall velocity $\vrc$,
two velocity dispersions of the infall component $\sr$ and $\st$, skewness
parameter $\alpha$, and kurtosis parameter $\dof$~(effectively reducing to
five parameters at $r>\rsh$). With seven parameters,
Equation~\ref{eqn:p2d} provides an excellent fit for the measured $\p2d$ at
all scales, as shown visually in the bottom panels of Fig.~\ref{fig:v2dfit},
and in greater detail below. We considered other parameterizations for the
infall component, such as sums of Gaussians, but we were unable to find a
compact description as accurate as the skewed $t$-distribution, so we
obtained poor results in modeling $\xicgs$.

\begin{figure*}
\centering
\resizebox{1.0\textwidth}{!}
{\includegraphics{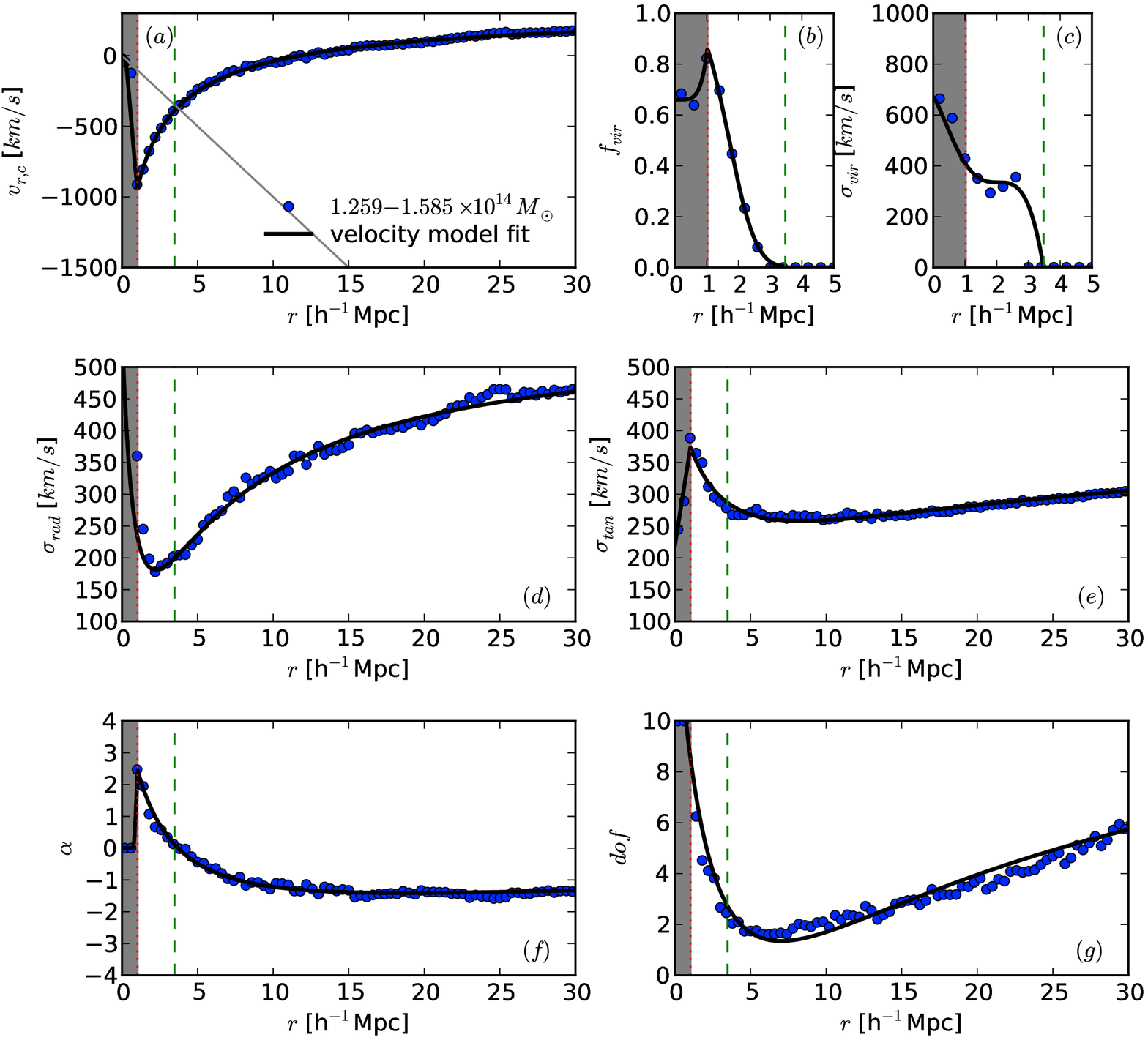}}
\caption{Functional fits~(black curves) to the radial profiles of
each GIK parameter measured in simulation~(blue circles). Similar
to Fig.~\ref{fig:parfit}, but only for the fiducial mass
bin~($M_c=1.259-1.585\times10^{14}M_\odot$). In each panel, the gray shaded
area below the radius of maximum infall indicates the regime where
cut--offs are required to match the mixing of virial and infall components
as shown in the left panel of Fig.~\ref{fig:v2dfit}. Dashed vertical lines
indicate the turn--around radius, where the Hubble flow~(gray solid line
in the top left panel) is equal to the characteristic infall velocity and
the skewness parameter $\alpha$ crosses zero in the bottom left panel.}
\label{fig:funfit}
\end{figure*}

Using the best--fit GIK models, we take a closer look into the properties
of $\p2d$ at different radii in Fig.~\ref{fig:pvcond}. In each panel,
the gray thick curve shows the 1D marginal probability distribution of
tangential~($P(v_t)$, top) or radial~($P(v_r)$, bottom) velocities. In the
left two columns where $r<\rsh$, the 1D marginal probability distributions of
$v_t$~(top) and $v_r$~(bottom) are decomposed into virialized~(red dashed) and
infall~(blue dotted) components. For the innermost bin, the infall component
has a much broader spread in radial velocities than in tangential velocities,
signaling its non-virial origin~(also see Fig.~\ref{fig:v2dfit}). The
second bin shows a more prominent infall component with much smaller radial
dispersion in the mixture~(bottom). In the right two columns where only
infall happens, we show the conditional probability distribution of $v_t$
at three fixed values of $v_r$ in the top panels, and vice versa in the
bottom panels. For the $r=4.0\--4.4\,\hmpc$ bin, as is also apparent in
Fig.~\ref{fig:v2dfit}, the conditional distributions show little skewness.
However, the conditional probability of $v_t$ shows higher kurtosis at more
probable values of $v_r$. In other words, if we divide the infall population
at fixed $r$ into streams of different $v_r$, the dominant streams have
a more sharply peaked tangential velocity distribution $P(v_t|v_r)$. For
instance in the 3rd column, the blue curve in the top panel shows $P(v_t|v_r)$
for the stream of $v_r=-480\,\mathrm{kms}^{-1}$, which is near the peak of
$P(v_r)$ according to the bottom panel. Compared to the other two streams
of zero~(green) and positive radial velocity~(red), the blue distribution
has many fewer galaxies with $|v_t| > 500\,\mathrm{kms}^{-1}$. Similarly,
the conditional probability of $v_r$ also shows higher kurtosis when $v_t$
is closer to zero~(e.g., green curve in the bottom panel).  This kurtosis
relation between $v_t$ and $v_r$ is also apparent in the $r=20.0\--20.4\,\hmpc$
bin, and it is ubiquitous at other distances. In addition, extra skewness
in $P(v_r)$ develops at other distances, and the degree of skewness
correlates with $v_t$, as seen in the bottom right panel. The skewness
switches sign at scales below the turn--around radius, where the radial
velocity distributions are negatively skewed~(second column, bottom panel,
blue dotted curve). The GIK model provides a faithful description of the
skewness and kurtosis measured in the simulations, but to avoid clutter,
we do not show the simulation measurements in Fig.~\ref{fig:pvcond} ---
the goodness of fit will be ultimately tested in the modeling of $\fvlos$.

\begin{figure*}
\centering
\resizebox{1.0\textwidth}{!}
{\includegraphics{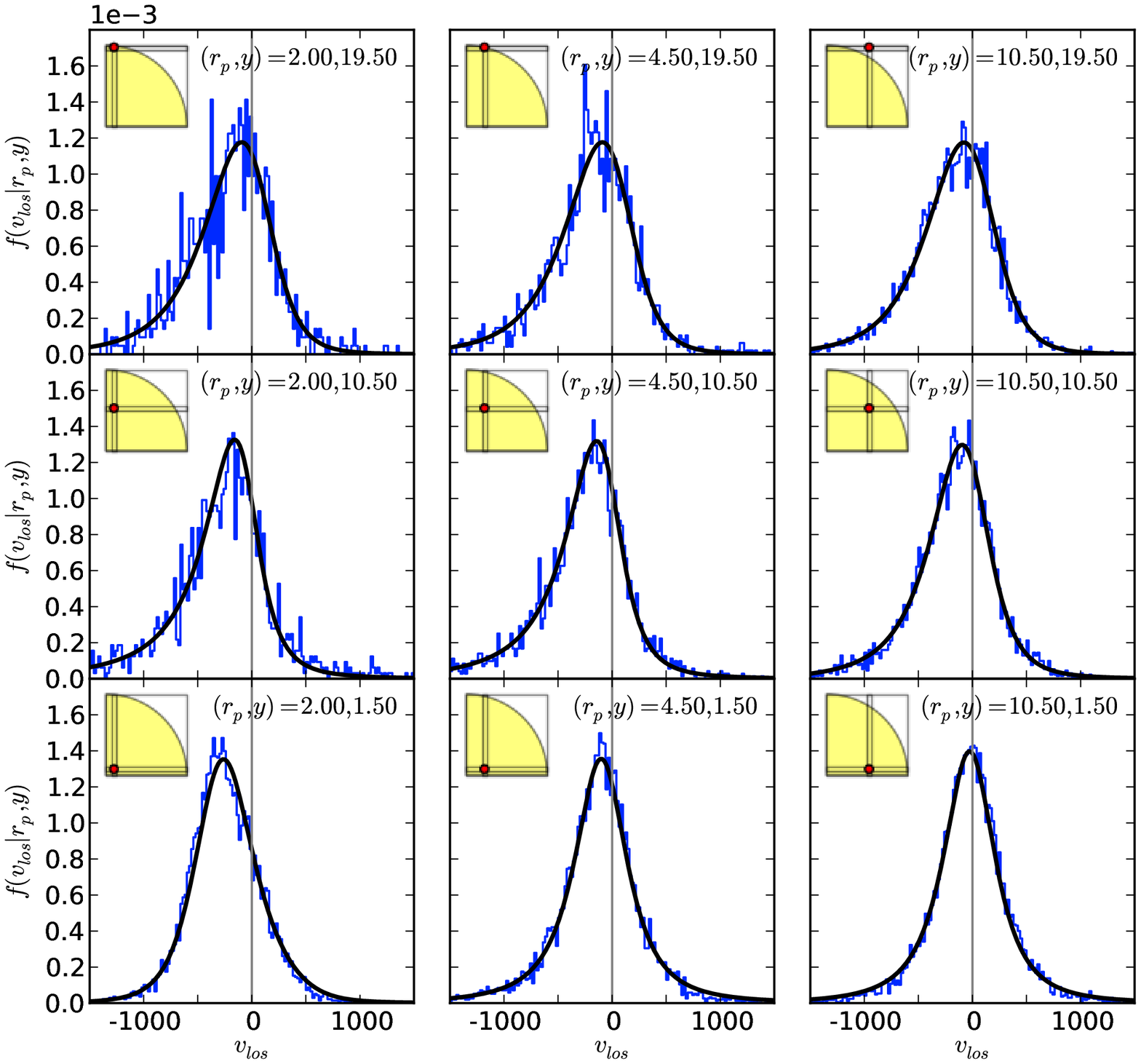}}
\caption{LOS velocity distribution in nine cells of different projected~($r_p$)
and LOS~($y$) separations in real space for the fiducial mass bin. Blue histograms and black curves show the measurements from
the simulation and the prediction from our best--fit GIK model. The inset panel
on top left of each panel indicates the relative position of each cell~(red
dot) relative to the cluster center~(yellow quadrant).}
\label{fig:vlospred}
\end{figure*}

\begin{figure*}
\centering
\resizebox{1.0\textwidth}{!}
{\includegraphics{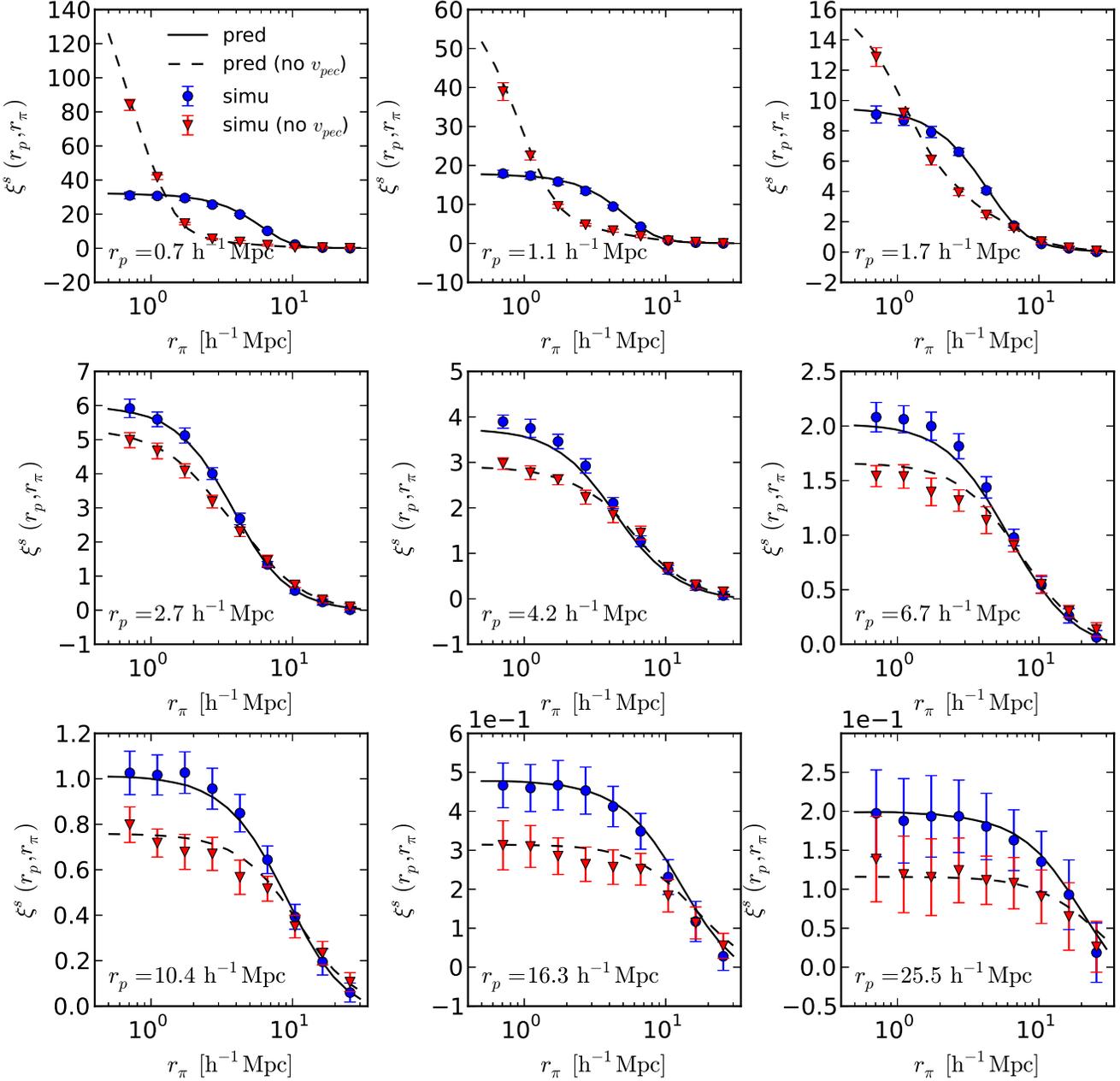}}
\caption{Comparison between the predicted and measured $\xicgs$ at nine different projected separations for the fiducial mass bin. Blue circles
and red triangles with error bars show the simulation measurements with
peculiar velocity turned on and off, respectively. Solid curves show the
predictions from the best--fit GIK model when peculiar velocity is on, while
dashed curves show the direction transformation from $\xicgr$ to $\xicgs$
when peculiar velocity is off.}
\label{fig:cutspred}
\end{figure*}

\begin{figure*}
\centering
\resizebox{0.80\textwidth}{!}
{\includegraphics{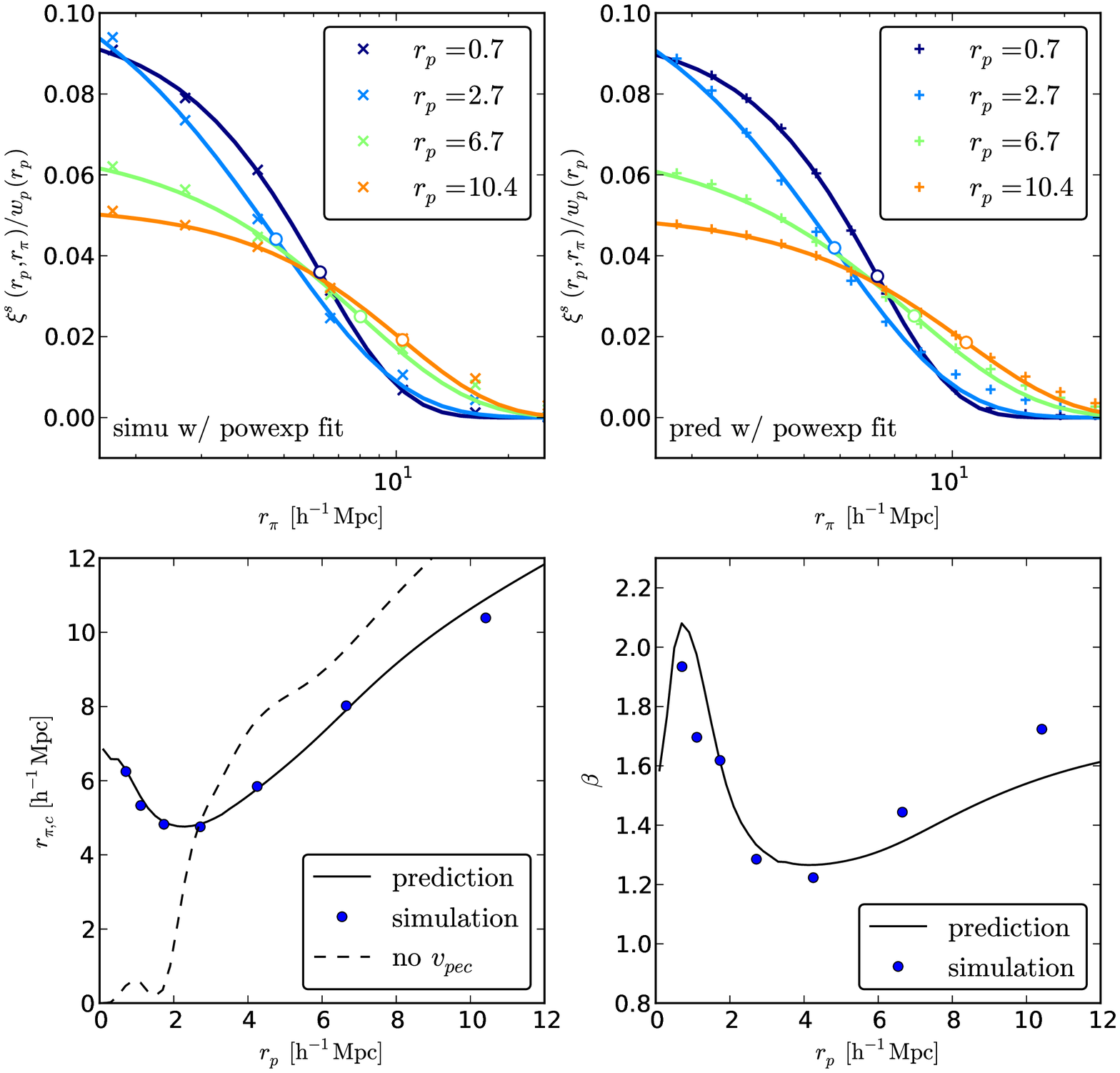}}
\caption{{\it Top panels}: Powered exponential fits~(solid curves) to the
measured~(top left) and predicted~(top right) profiles of $\xicgs/w_p$
at different $r_p$ for the fiducial mass bin.  {\it Bottom
panels}: Best--fit characteristic LOS distance $r_{\pi, c}$~(bottom left) and
shape parameter $\beta$~(bottom right) from the top panels; Blue circles and
solid curves show the best--fits to simulation measurements and predictions
from the GIK model, respectively. The dashed curve in the bottom right panel shows
the best--fit $r_{\pi, c}$ for the $\xicgs$ measured when peculiar velocity is turned off.}
\label{fig:ushape}
\end{figure*}

We fit the GIK model with seven parameters to the measurements of $\p2d$
at radial bins from $0$ to $40$ $\hmpc$ for six bins of clusters in the
simulation.  The results are shown in Fig.~\ref{fig:parfit}, and we will
discuss each panel in turn.  The top left panel shows the profiles of the
characteristic infall velocity $\vrc$. At given mass, the absolute value of
$\vrc$ becomes larger with decreasing distance, peaking at some characteristic
radius of maximum infall~$\rmi$. Below $\rmi$, as we see in the left column
of Fig.~\ref{fig:v2dfit}, the infall component blends into the virialized
population, reducing $\vrc$ sharply to zero. The amplitude of $\vrc$ scales
with mass as approximately $M^{1/3}$, therefore providing a clear diagnostic of
cluster masses. The top middle and right panels show the profiles of virialized
fraction $\f$ and dispersion of the virial velocities $\sv$, respectively. The
cutoffs below $\sim 1\,\hmpc$ are caused by the same blending effect below
$\rmi$, where $\f$ stays approximately constant at $0.65$. In this regime,
the seperation between the virialized and infall components is no longer
sharp~(see Fig.~\ref{fig:v2dfit}, left), so while our fit to $\p2d$
is accurate, the physical significance of individual parameters is less clear.
There are plateaus in the $\sv$ profiles at $r$$\sim$$2\--4\,\hmpc$ depending
on mass, possibly indicating the pre--heating induced by shear flow at the
surface where infall and virialized component first contact; however, the
virialized component is only $\sim10\%$ of the total at these radii. The
amplitude of $\sv$ profiles and the extents of both $\sv$ and $\f$ profiles
scale with mass. Returning to the infall component, cluster masses affect both
the amplitude and shape of the $\sr$~(middle left) and $\st$~(middle right)
profiles.  More massive clusters induce infall streams with larger~(smaller)
$\sr$ on smaller~(larger) scales, while $\st$ increases monotonically with
cluster mass on all scales $r\leq20\,\hmpc$.  Finally, there is little
variation of the profiles of $\alpha$ and $\dof$ with mass. The $\alpha$
profiles cross zero at different distances depending on the $\vrc$ profile
of each mass bin, but since $\alpha$ decreases slowly with distance at fixed
mass, the amplitudes of different curves do not change much.  The degrees
of freedom start from extremely high values~(near--Gaussian tails) at the
cluster center, reach a minimum at $\sim5\,\hmpc$~(a Lorentz distribution has
$\dof=1$), then rise up again on large scales. Despite the insensitivity to
cluster mass, the systematic variation of $\alpha$ and $\dof$ with radius,
not otherwise captured by simple Gaussian or exponential streaming models,
is pivotal to the accurate modeling of $\fvlos$ and $\xicgs$ at relevant
scales. The interdependence of radial and tangential velocities, seen in
Fig.~\ref{fig:pvcond}, are aslo required for accurate modeling.  The cutoffs
of $\st$ and $\alpha$ profiles on small scales have the same blending origin
as those of $\vrc$ and $\f$.

\begin{figure*}
\centering
\resizebox{0.80\textwidth}{!}
{\includegraphics{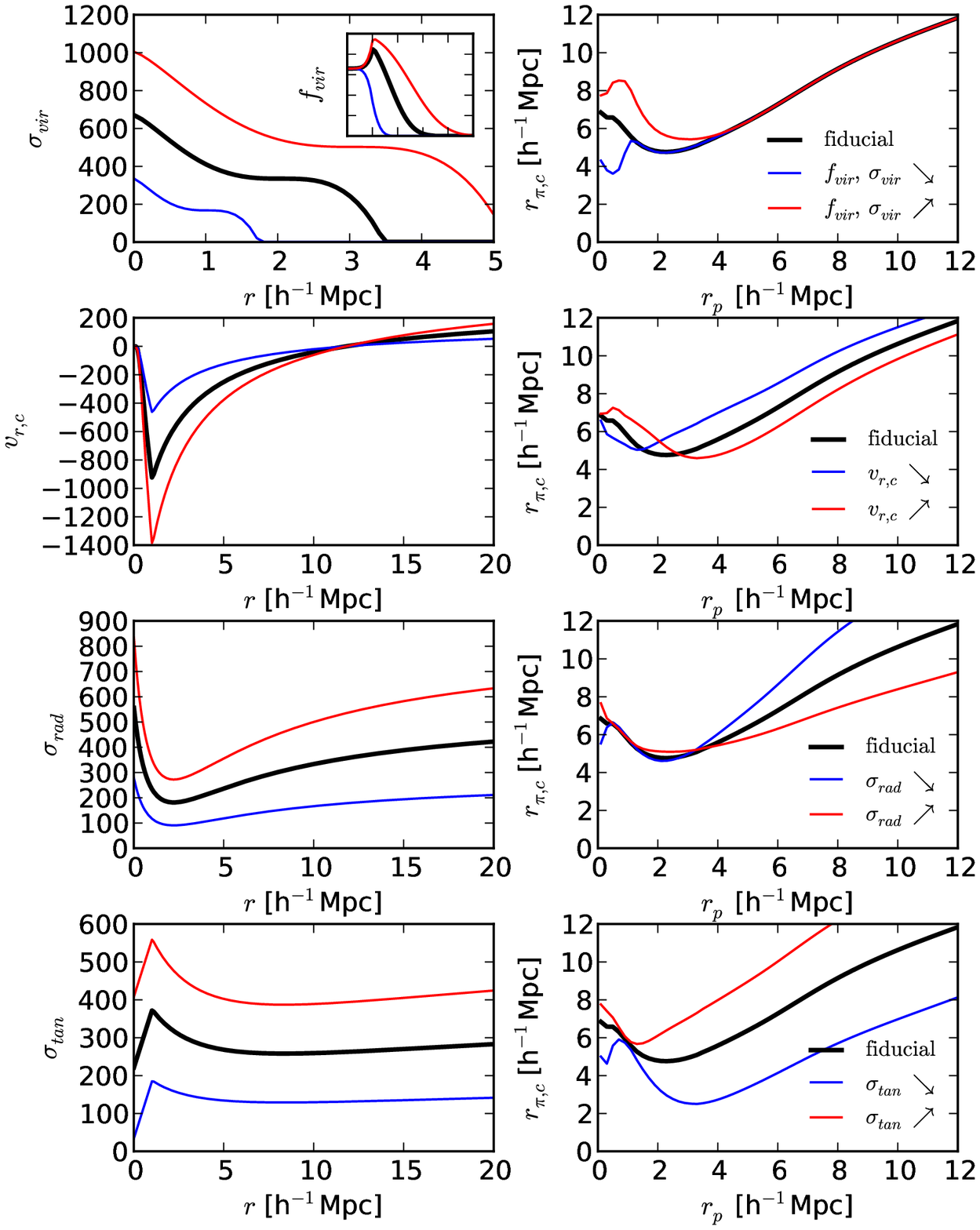}}
\caption{Effects of perturbing each GIK component on the characteristic U-shaped
curve of $\xicgs$. In each row, the amplitude of one of the components~(from
top to bottom: fraction and velocity dispersion of the virialized component
$f_{vir}$ and $\sigma_{vir}$ , radial velocity profile $v_{rad}$, radial
velocity dispersion $\sigma_{rad}$, and tangential velocity
dispersion $\sigma_{tan}$) is changed from its fiducial value by $\pm 50\%$, and
we fit powered exponential functions to the predicted $\xicgs$
at each $r_p$. The right panels show the impact of these changes on the $r_{\pi,
    c}(r_p)$ curves, comparing the predictions of the modified model~(blue and
red) to the fiducial model~(black).}
\label{fig:perturb}
\end{figure*}

To summarize, Fig.~\ref{fig:parfit} shows that higher mass clusters have,
as expected, higher amplitude characteristic infall curves $\vrc$, virialized
components with higher velocity dispersion $\sv$ that extend to large radii
$\rsh$, and higher tangential velocity dispersions $\st$ within the infall
component. The radial dispersion $\sr$ has weaker mass dependence that reverses
sign at $r$$\sim$$5\,\hmpc$, and the profiles of $\alpha(r)$ and $\dof(r)$,
which control the shape of the distribution function of the infall component,
show nearly universal, mass--independent bahavior. We are especially interested
in extracting the characteristic infall curves $\vrc(r)$, as these probe the
extended mass profile around clusters and should be insensitive to physics
that may alter the dispersions of satellite galaxies within halos. The
smooth behaviour and systematic trends in Fig.~\ref{fig:parfit} suggest that
isolating $\vrc$ will be feasible, and the profiles of $\sv(r)$ and $\st(r)$
offer additional mass diagnostics. The main area of uncertainty is the cutoff
behaviour below the maximum infall radius $\rmi\approx1\,\hmpc$. This should
have little impact in practical applications, as shot noise and ``fiber
collision'' effects~\citep{blanton2003} make $\xicgs$ difficult to measure
at small $r_p$, and $\xicgs$ at a given $r_p$ is absolutely unaffected by
any scales $r<r_p$.

\begin{figure*}
\centering
\resizebox{0.9\textwidth}{!}
{\includegraphics{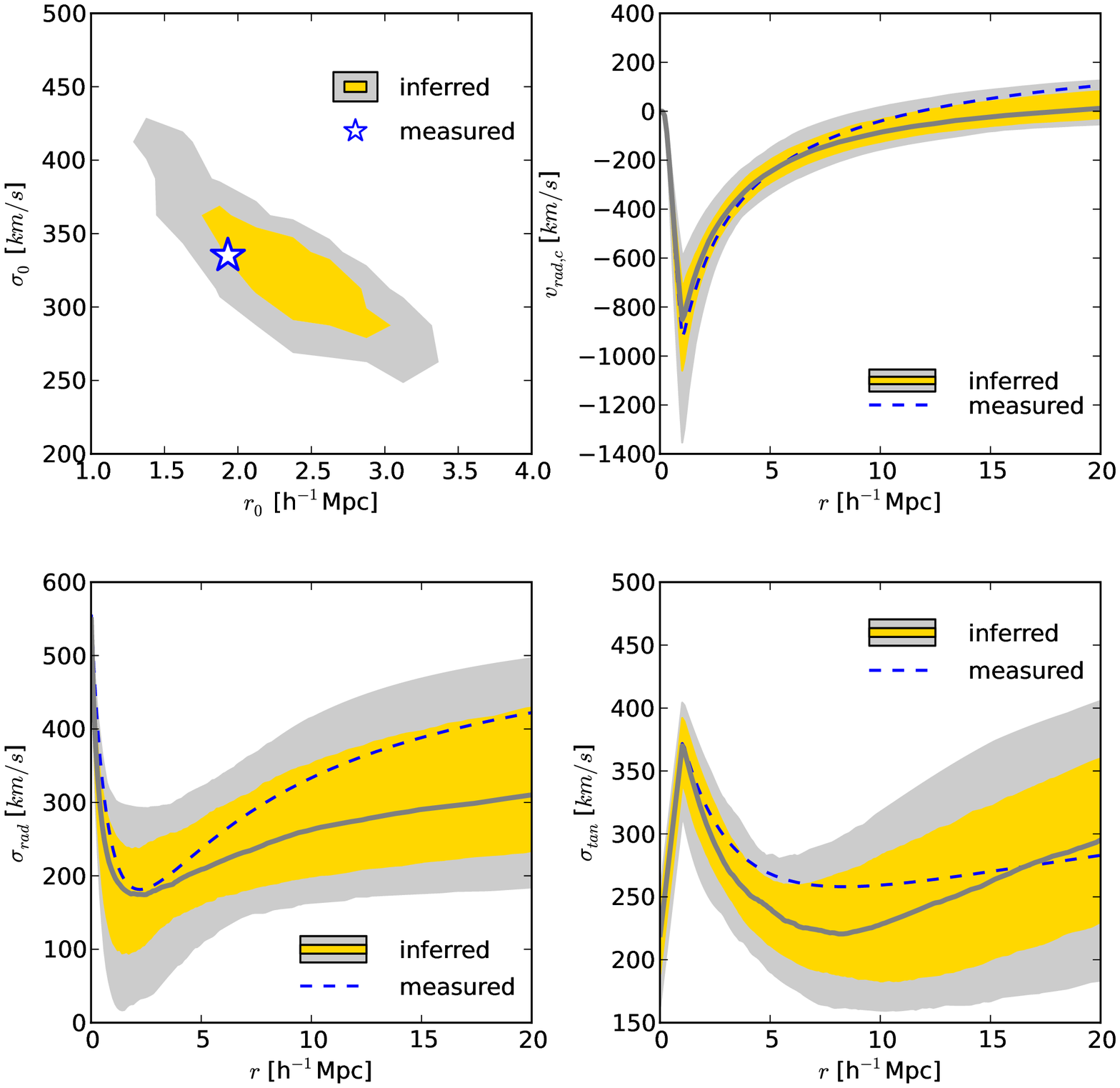}}
\caption{Constraints on each component of the GIK model from the simulation
measurements of $\xicgs$. Yellow and gray contours indicate the $68
\%$ and $95\%$ confidence regions. The blue star in the top left panel and
dashed curves in the remaining panels show the direct measurements from the simulation.}
\label{fig:mcmcmill}
\end{figure*}

To describe the variations of these seven parameters with $r$, we fit them
with analytic functions chosen to match the numerically measured shapes. We
fit $\vrc(r)$ with a 3-parameter function,
\begin{equation}
\vrc(r) = q_1 - \frac{p_1}{(r+\rmi)^{\beta_1}},
\label{eqn:vrc}
\end{equation}
where $\rmi$ is the maximum infall radius beyond which $\vrc$ is exponentially
cut off to zero, $q_1$ controls the maximum amplitude, and $p_1$
and $\beta_1$ together control the asymptote and slope of the large scale
power--law behaviour, respectively. The virialized fractions are well described by
a ``powered exponential'',
\begin{equation}
\f(r) = \exp\left\{-\left(\frac{r}{r_0}\right)^3\right\},
\label{eqn:fvir}
\end{equation}
where $r_0$ is a characteristic radius of the cluster. We set the radius at
which $\f(r)$ falls to $0.3\%$ as the shock radius $\rsh$, which corresponds
to $1.8\,r_0$, and we set $\f(r>\rsh)=0$. The velocity dispersion profiles of
the virialized component can be described by
\begin{equation}
\sv(r) = \s0 \left\{1 + \left( 1 - \frac{r}{r_*}\right)^\nu \right\}^3, 
\label{eqn:sv}
\end{equation}
where we find $\nu=1.315$ provides a good fit to all the mass bins, and we
impose the constraint that the plateau happens at $r=\rsh$, so that $r_*$ is
determined by the best--fit $\rsh$ from $\f$ via $r_*\equiv 
2^{-1/\nu}\rsh=0.59\,\rsh$. The remaining four nuisance profiles are well described by
one generic function,
\begin{equation}
\left\{\sr,\,\st,\,\alpha,\,\dof\right\} (r) = q_{i} -
p_{i}\frac{r}{(r+r_{i})^{\beta_{i}}},\quad\mbox{$i \in \{2,3,4,5\}$},
\label{eqn:pqseries}
\end{equation}
where the effects of $q_i$, $p_i$, and $\beta_i$ are similar to that of $q_0$,
$p_0$, and $\beta_0$ in Equation~\ref{eqn:vrc}, except that the minima happen
at $r_i/(\beta_i-1)$ instead of $r_i$.

The best--fit functions are shown in Fig.~\ref{fig:funfit} for the fiducial
mass bin. The gray shaded area in each panel indicates the $r<\rmi$ regime,
where we apply simple cut--offs~(exponential or polynomial) to mimic the
rough measurements from simulation. We emphasize again that the cut--offs
do not affect the modeling of $\fvlos$ and $\xicgs$ at $r_p>\rmi$. The gray
line in the top left panel indicates the amplitude of Hubble flow, crossing
$\vrc(r)$ at the turn--around radius, which is indicated in each panel by
the vertical dashed line.

\section{Modeling of $\mathbf{\xicgs}$}
\label{sec:xicgs}

The GIK model describes average galaxy motions around clusters in 3D, but
the modeling of $\xicgs$ requires predicting the 1D galaxy motions projected
along the LOS at any given ($r_p, y$). Fig.~\ref{fig:vlospred} compares the
LOS velocity distribution $\fvlos$ measured directly from the simulation
to that predicted from the best--fit GIK model~(i.e., using the functional
fits shown in Fig.~\ref{fig:funfit}), at nine different cells around clusters
in the fiducial mass bin. The relative position of each cell around the cluster
center is indicated by the red dot in the inset quadrant. Although the GIK model is calibrated
using the same simulation that $\fvlos$ is measured from, the agreement we
see in Fig.~\ref{fig:vlospred} in all panels is highly non--trivial ---
the 2D mixture model accurately recovers the varying degrees of skewness
and kurtosis of LOS velocity distributions at different positions of $(r_p,
y)$, whereas a less comprehensive model~(e.g., only fitting to the first and
second moments, or treating tangential and radial velocities independently)
would fail. The agreement in Fig.~\ref{fig:vlospred} also reconfirms that our
best--fit GIK model provides an excellent description of the galaxy kinematics
in the Millennium simulation.

Once we predict $\fvlos$, it is straightforward to predict $\xicgs$ by
convolving $\fvlos$ with the real--space cluster--galaxy correlation
function $\xicgr$, then compare to the $\xicgs$ measured directly from
simulation. The convolution is, expressed in a form more explicit than
Equation~\ref{eqn:convolution},
\begin{eqnarray}
\xicgs(r_p, r_\pi) +1 &=& H_0 \int_{-\infty}^{+\infty}
\left[\xicgr\big(\sqrt{r_p^2+y^2}\big)
+ 1\right] \times \nonumber\\
&&f\big(H_0 (r_\pi - y)|r_p, y\big) d y.
\label{eqn:convolution2}
\end{eqnarray}
To make sure that the $\xicgs$ comparison is unaffected by any inaccuracies
in $\xicgr$, we use the $\xicgr$ directly measured from simulation to convolve
with $\fvlos$.  For measuring $\xicgr$, we count the numbers of galaxies around
clusters in spherical shells of successive radii, ranging from $10\,\hkpc$
to $50\,\hmpc$ with logarithmic intervals, average over all clusters in each
bin, and normalize by the galaxy numbers expected in a randomly located shell
of equal volume. We measure $\xicgs$ in
a similar way, counting galaxies in cylindrical rings of successive
$r_\pi$ for each $r_p$~(assuming a distant--observer approximation so that the
LOS is an axis of the box).

Fig.~\ref{fig:cutspred} compares the $\xicgs$ from convolution to the
simulation measurements for the fiducial mass bin from $r_p=0.7\,\hmpc$
to $25.5\,\hmpc$~($r_p$ increasing from top to bottom, left to right).
In each panel, blue circles with errorbars are the $\xicgs$ measured
directly from the simulation, while the solid curves are predictions from
the best--fit GIK model using Equation~\ref{eqn:convolution2}; red triangles
with errorbars are the $\xicgs$ measured when the peculiar velocity of
each object is set to zero in the simulation, i.e., setting $\xicgs(r_p,
r_\pi)\equiv\xicgr\left(\sqrt{r_p^2+r_\pi^2}\right)$. There is overall
good agreement between $\xicgs$ predicted from the best--fit GIK model and
measured from simulation, except for the $r_p=4.2\,\hmpc$ and $6.7\,\hmpc$
panels where the model predictions are slightly higher than the measurements
at small $r_\pi$. The minor discrepancy probably comes from the stochastic
noise in measurements along one particular sight line~(similar to cosmic
variance), as there is already discrepancy for the no--$v_\mathrm{pec}$
cases~(red triangles vs.  dashed curves) in the same two panels before
convolving with $\fvlos$ --- the redshift space correlations are measured
along the $z$--axis of simulation box, while the real space correlation is
measured assuming isotropy of the entire box.

By comparing the two cases~($v_\mathrm{pec}$ vs. no--$v_\mathrm{pec}$) within each panel,
Fig.~\ref{fig:cutspred} also nicely illustrates the two major RSD effects ---
$\xicgs$ along the LOS is suppressed by random dispersion for small $r_p$,
but is amplified by galaxy infall for large $r_p$. To further quantify these
deformations, we fit a powered exponential function to the $\xicgs$ at each
$r_p$, which is equivalent to fitting a normalized powered exponential function
to $\Xi$ 
\begin{equation}
\Xi(r_\pi|r_p) \equiv \frac{\xicgs(r_p, r_\pi)}{w_p(r_p)}\bigg|_{r_p} 
\sim \exp \left\{-\left|\frac{r_\pi}{\rpic}\right|^\beta\right\},
\end{equation}
where $\rpic$ is the characteristic length scale at which $\xicgs$ drops to
$1/e$ of its maximum value at $r_\pi=0$. The shape parameter $\beta$ yields a
Gaussian cutoff for $\beta=2$ and simple exponential for $\beta=1$, though any
value is allowed in the fit.

Fig.~\ref{fig:ushape} summarizes the fits of $\Xi$ measured from
simulation~(crosses in the top left panel) and predicted by the best--fit
GIK model~(plus symbols in the top right panel) at four different $r_p$. The
fits~(solid curves) in the two panels are very similar, so we focus on
the top left panel here.  The open circle through each curve indicates the
position of $\rpic$, which migrates from $\sim6.5\,\hmpc$ at $r_p=0.7\,\hmpc$
inward to $\sim5.0\,\hmpc$ at $r_p=2.7\,\hmpc$, and then outward to $\sim
8$ and $10\,\hmpc$ at $r_p=6.7$ and $10.4\,\hmpc$, respectively. This
``precession'' of $\rpic(r_p)$ is more clearly shown in the bottom left
panel, where the blue circles and solid curves plot the migration of $\rpic$
as function of $r_p$ from fits to the measurements and model predictions,
respectively. This characteristic U-shaped curve is the same one shown in the
right panel of Fig.~\ref{fig:xirsdemo}.  Also shown in the bottom left panel is
$\rpic(r_p)$ for the no--$v_\mathrm{pec}$ case, which grows monotonically with
increasing $r_p$. This can be easily understood for a power--law $\xicgr\propto
r^{-\gamma}$, where $\xicgs(r_p, r_\pi)\propto(1+(r_\pi/r_p)^2)^{-\gamma/2}$
with no $v_\mathrm{pec}$, hence $\rpic\propto r_p$. The change of slope
around $r_p\simeq 2.5\,\hmpc$ for the no--$v_\mathrm{pec}$ case is caused
by the change of $\gamma$ during transition from the 1-halo to 2-halo
regime. The bottom right panel shows the corresponding changes of $\beta$
as function of $r_p$, which largely follow~(in a slightly lagged fashion)
the variations in $\rpic(r_p)$, and which are again well described by the
GIK model.  We will focus on $\rpic(r_p)$ as the representative feature of
$\xicgs$ in the next section.

Note that the powered--exponential is not intended to be a viable model
of $\Xi$, just a compact and visually appealing way of quantifying the
characteristics of $\Xi$. Therefore, although the powered--exponentials do
not fit well on scales larger than $\sim 15\,\hmpc$, the best--fit $\rpic$
and $\beta$ still effectively capture the main features of each curve.

\section{Bayesian Inference of Velocity Distribution}
\label{sec:mcmcmill}

Armed with an accurate GIK model that correctly predicts the $\xicgs$
signal, we are in the position to investigate the origin of the ``U-shaped
curve'' of Fig.~\ref{fig:xirsdemo}, and, more importantly, to examine the
intrinsic degeneracies within the modeling of $\xicgs$, which carries valuable
information that we hope to exploit robustly.  To achieve this understanding,
we perturb the elements of the velocity field one at a time, changing the
amplitude of a single model
component by $\pm 50\%$ while holding others fixed. In \S\ref{sec:gik} we found that among the seven parameters in the GIK model~(see Fig.~\ref{fig:parfit}),
$\alpha$ and $\dof$ are insensitive to cluster mass, so we focus on the
remaining five parameters, linking $\sv$ and $\f$ so that there are four
independent GIK components. Each row in Fig.~\ref{fig:perturb} shows the
results of perturbing the amplitude of one component: properties of the
virialized population $\sv$+$\f$, characteristic infall velocity $\vrc$, radial
velocity dispersion of the infall population $\sr$, or tangential velocity
dispersion of the infall population $\st$. For the first row, we change the
amplitude of $\sv$ by changing the value of $\s0$ in Equation~\ref{eqn:sv},
and we simultaneously change the value of $\rsh$ proportionally with $\s0$,
which in turn expands $\f$ along the $r$ axis~(the inset axis). For the rest,
since they all have similar parameterization, we change the amplitudes via
multiplying $p_i$ and $q_i$ in Equation~\ref{eqn:vrc} and~\ref{eqn:pqseries}
by the same factor. We will describe each row in turn, from top to bottom.
\begin{itemize}
\item When $\sv$ is increased and $\f$ is expanded, the virialized
region becomes hotter and larger, producing more flattened $\xicgs$ for
$r_p<\rsh$ while having no effect for $r_p\geq\rsh$. This is the portion of
the FOG effect caused by virialized dispersion.
\item When the infall velocity $\vrc$ is faster~(more negative), galaxies at
    large $r_p$ are shifted closer to the cluster center~(more ``Kaiser
    compression''), therefore
reducing $\rpic$. However, at small $r_p$ the infall is strong enough to send
galaxies from one side of the cluster in real
space~($y<0$ or $y>0$) to the opposite side in redshift space~($r_\pi>0$
or $r_\pi<0$). This is the portion of the FOG effect caused by infall. The characteristic
projected separation $r_p^{*}$ at which the U-shaped curve reaches a minimum
is then set by the transition from large--scale compression to small--scale
inversion, shifting to a larger scale when infall becomes stronger. The value of
$\rpic$ at the minimum also decreases slightly as $\vrc$ increases.
\item When $\sr$ is higher, the velocity ellipses of $\p2d$ are more
elongated along the radial velocity axis. Since the ellipses sit mostly at the
negative half of the radial velocity axis, higher $\sr$ effectively leads to
overall stronger infall. However, at smaller LOS distance~($y < r_p$), the
LOS velocity distribution is insensitive to the changes in $\sr$~($\theta <
\pi/4$ in Equation~\ref{eqn:fvlos}), so the stronger infall only starts
to affect $\rpic$ at large scales where $\rpic$ is comparable to $r_p$,
and the impact on $\rpic$ increases as a function of $r_p$.
\item When $\st$ is higher, the velocity ellipses of $\p2d$ are
more stretched along the tangential velocity axis, effectively increasing the
dispersions of $v_\mathrm{los}$ without modifying the mean. Therefore, similar to the
effect of $\sv$ on small scales, higher $\st$ increases $\rpic$ on all scales,
though the fractional impact is largest at $r_\pi\gtrsim\rsh$.
\end{itemize}
The basic U-shape of the $\rpic$ vs. $r_p$ curve is straightforward to
understand: FOG stretching at small $r_p$ gives way to Kaiser compression at
intermediate $r_p$ which gives way to Hubble flow expansion at large
$r_p$.\footnote{Even at large $r_p$, infall reduces $\rpic$ relative to the real
space value.} However, Fig.~\ref{fig:perturb} shows that the detailed shape of
this curve, and more generally of $\xicgs(r_p, r_\pi)$, reflects a complex
interplay among the four components of the galaxy kinematics around clusters.

\begin{figure*}
\centering
\resizebox{1.0\textwidth}{!}
{\includegraphics{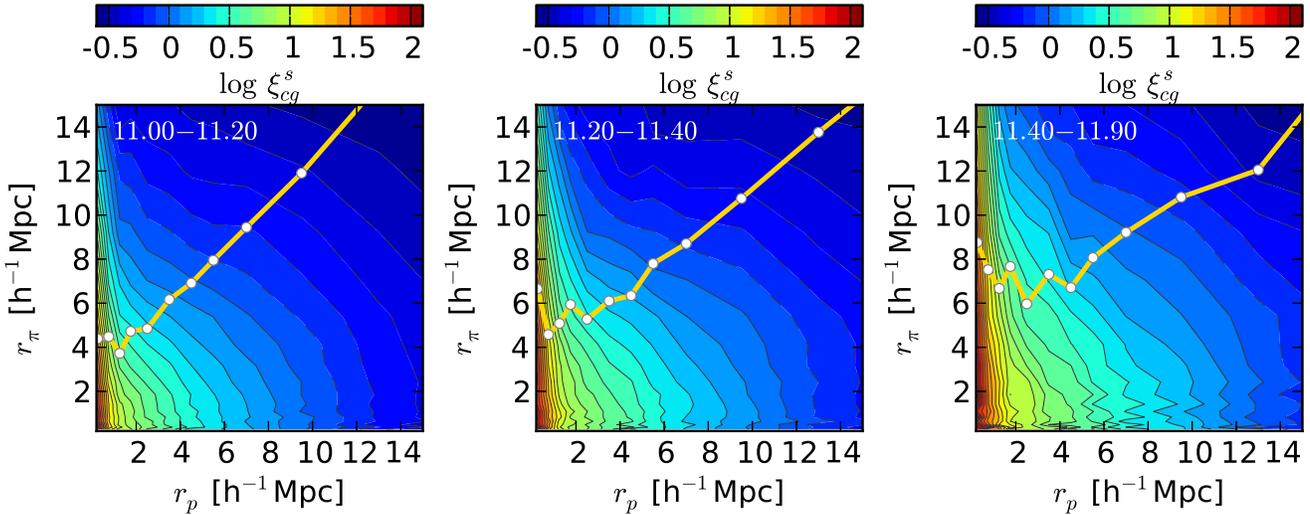}}
\caption{Measurements of $\xicgs$ for SDSS groups in three bins of
BCG stellar mass. The yellow U-shaped curve in each panel shows the characteristic
LOS distance $r_{\pi,c}$ at each projected separation $r_p$. The contour
scales are the same for all panels,  indicated by the colour bars on top.}
\label{fig:xicgsdss}
\end{figure*}

Crucially, the impact of each GIK component has a distinct scale and amplitude
dependence, suggesting that they can be inferred from $\xicgs$ measurements with
only limited degeneracy. 
To confirm this expectation, we construct a Gaussian likelihood model with the
measurements of $\xicgs$ in Fig.~\ref{fig:cutspred} as the input
data, and the functional parameters introduced in \S\ref{sec:gik}
as our model parameters. The parameters we vary in the
model are, $\{q_1$, $p_1$, $\beta_1\}$ for the characteristic infall
velocity~(Equation~\ref{eqn:vrc}), $\{r_0$, $\s0\}$ for the virialized
component~(Equations~\ref{eqn:sv} and~\ref{eqn:fvir}), $\{q_2$, $p_2$, $r_2$,
$\beta_2\}$ for the radial velocity dispersion~(Equation~\ref{eqn:pqseries}),
and $\{q_3$, $p_3$, $r_3$, $\beta_3\}$ for the tangential velocity
dispersion~(Equation~\ref{eqn:pqseries}). In this way, we allow maximum
freedom for the shape and amplitude of each component to vary during
the fit.  We keep the remaining parameters fixed to their best--fit
values in Fig.~\ref{fig:funfit}, as they appear to be insensitive to cluster
masses~(see Fig.~\ref{fig:parfit}) --- we keep the radial profiles of $\alpha$
and $\dof$ fixed in amplitude and shape, and we maintain the shapes of the $\sv$
and $\f$ profiles, while allowing them to change scale along both axes via $r_0$
and $\s0$. The Gaussian log--likelihood is thus
\begin{equation}
\ln \mathcal{L}(\xicgs|\boldsymbol{\theta}) \propto 
-\frac{1}{2}
\left(\xicgs-\xicgs(\boldsymbol{\theta})  \right)^T C^{-1}
\left(\xicgs-\xicgs(\boldsymbol{\theta})  \right),
\label{eqn:lnl}
\end{equation}
where 
\begin{equation}
\boldsymbol{\theta} \equiv \{q_1, p_1, \beta_1, r_0, \s0, q_2, p_2, r_2,
\beta_2, q_3, p_3, r_3, \beta_3 \},
\label{eqn:varying}
\end{equation}
and $C$ is the data covariance matrix measured from Jackknife re--sampling of
the simulation volume.\footnote{We divide the simulation box into octants, and
    derive the covariance from the $\xicgs$ measurements of the whole box
    and eight subsamples, each composed of the whole box minus one octant.}
For demonstration purposes, we use the direct simulation measurements of
$\xicgr$ to convolve with the predicted $\fvlos$. When applying the model to
observations, $\xicgr$ should either be inverted from the measured $w_p$~(see
\S\ref{sec:mcmcsdss}) or directly predicted from theoretical models~(papers
II and III).

We adopt a Bayesian approach, assuming uninformative priors for all the
parameters that are allowed to vary. For the parameter inference, we sample
the posterior distributions using a Markov Chain Monte Carlo~(MCMC), where an
adaptive Metropolis step method is utilized during the burn--in period. The
whole chain has 15,000 iterations, 5,000 of which belong to the burn--in
period, where auto--correlation tests show good convergence before
actual sampling. We emphasize that while we used the $\rpic(r_p)$ curve for
illustration in Fig.~\ref{fig:perturb}, our statistical inference of GIK uses
the full $\xicgs(r_p, r_\pi)$, as indicated by Equation~\ref{eqn:lnl}.

Fig.~\ref{fig:mcmcmill} presents the constraints on the four GIK components
inferred from the MCMC when we apply this fitting procedure to the $\xicgs$
measurements from the fiducial mass bin. Yellow and gray contours show the $68\%$ and $95\%$
confidence limits, respectively. 
For the constraints on $\vrc$, $\sr$, and $\st$ profiles, we compute the median~(68\%
and 95\%) curves discretely at each $r$ as the median~(68\% and 95\%) functional
values calculated from all the iterations at that $r$, rather than the
functional values calculated from the median~(68\% and 95\%) parameters.
The blue star in the top left panel and blue dashed curves in the other
panels indicate the direct fits to simulation measurements of GIK in
Fig.~\ref{fig:funfit}. The constraints derived from $\xicgs$ show overall
agreement with the direct fits, confirming that the distinctive effect of
each GIK component allows us to constrain the overall model with minimal
ambiguity. In particular, the characteristic radial infall curve $\vrc(r)$
is inferred with small uncertainty from this cluster
sample~($N_\mathrm{cluster}=691$) .  The medians
are slightly offset from the fiducial values, possibly because of the same
stochastic noise that causes the discrepancies in Fig.~\ref{fig:cutspred}, but
the offsets are smaller than the $68\%$ statistical uncertainty. 

\section{Application to SDSS Groups}
\label{sec:mcmcsdss}

We defer a comprehensive calibration and application of the GIK model to
the future, but as a proof of concept, we apply the current GIK model to
$\xicgs$ measured for rich galaxy groups found in the SDSS. We employ the
group catalog of~\cite{yang2007} in the local universe~($z<0.2$) found in the
spectroscopic data of SDSS Data Release 7~\citep[DR7;][]{abazajian2009}. Each
group was identified initially with a friends-of-friends scheme in the
redshift space and kept in the catalog by an iterative adaptive filter
method. For more details on the construction of the catalog, we refer
the reader to~\cite{yang2007}. Small groups are likely to suffer more
contamination and to have weaker infall patterns. We therefore select $6691$
groups that have relatively high stellar mass~($\log M_*/M_\odot\ge11.0$)
in their brightest cluster galaxies~(BCGs) and have at least three identified
galaxy members; We describe these groups as our ``cluster'' sample. We divide
the $6691$ clusters further into three BCG stellar mass bins, with $\log
M_*=11.00\--11.20$~($N_{c}=4018$), $\log M_*=11.20\--11.40$~($N_{c}=2027$),
and $\log M_*=11.40\--11.90$~($N_{c}=646$), respectively. For the first--cut
analysis here, we will use the positions of BCGs as the cluster centers,
without modeling the mis--centering effect~\citep{skibba2011}. For the galaxy
sample, we use the {\tt dr72safe0} sample within the NYU Value-Added Galaxy
Catalog~\citep{blanton2005} derived from the main spectroscopic sample in
DR7, containing $N_g=534206$ galaxies with K+E--corrected $r$-band magnitudes
between $-24$ and $-16$ at $z<0.2$.

In contrast to the simulation where the expected number of cluster--random
galaxy pairs is known precisely, we have to construct random catalogs of
galaxy samples that have the same angular and redshift selection functions as
the observed galaxy sample. We generate the angular completeness map in the
format of non--overlapping polygons from the {\tt dr72safe0} window function
using MANGLE~\citep{swanson2008}, then draw $N_g$ random galaxy coordinates
from each polygon based on its spectroscopic completeness. To account for the
redshift selection function, we randomly shuffle the redshifts of observed
galaxies and assign them to the random coordinates. This shuffling procedure
is equivalent to drawing redshifts from the parent redshift distribution
of the observed sample when $N_g$ is large. To reduce the random noise,
we repeat the process for ten times and construct a random galaxy sample
with $N_r=10\times N_g$.

We measure $\xicgs$ using the~\cite{davis1983} estimator,
\begin{equation}
\xicgs(r_p,r_\pi) = \frac{N_r}{N_g}\frac{N_{CG}(rp, r_\pi)}{N_{CR}(rp, r_\pi)} -
1,
\end{equation}
where $N_{CG}$ and $N_{CR}$ are the cluster--galaxy pairs and cluster--random
galaxy pairs, respectively. We estimate measurement uncertainties on $\xicgs$
using the Jackknife re--sampling method.\footnote{Since the number of clusters
in each bin is small compared to
    galaxies, we construct Jackknife sub--samples by dropping individual
    isolated clusters or groups of angularly close clusters from
each bin. The number of such sub--samples for each bin is $>400$.} In
addition, we also measure the projected cluster--galaxy correlations $w_p$
from the same data and random galaxy samples using an integration length
of $r_\pi^\mathrm{max}=40\,\hmpc$.

Fig.~\ref{fig:xicgsdss} shows the $\xicgs$ measured for the three bins of
clusters and the characteristic U-shaped curves derived from $\xicgs$. The
colour scales are the same in all panels, so it is clear that clusters in higher
stellar mass bins have on average higher masses, showing stronger correlation
signals at fixed ($r_p, r_\pi$). The BCG stellar mass, however, is only a loose
indicator of total cluster mass. To completely model $\xicgs$
measured in bins of BCG stellar mass, we should predict $\xicgr$
and $\fvlos$ as functions of $M$ to get $\xicgs(M)$, then 
convolve $\xicgs(M)$ with scatter in the $M_*$--$M$ relation weighted by the 
cluster mass function $dn/dM$. This type of comprehensive model has been used to
interpret weak lensing measurements~\citep{sheldon2009} for SDSS MaxBCG clusters
by~\cite{rozo2010},~\cite{tinker2012}, and~\cite{zu2012}, and we will adopt it
for $\xicgs$ modeling in future work. For our first--cut analysis here, we will
infer {\it average} GIK properties of the~\cite{yang2007} groups by treating
each $M_*$-bin as though it were a single halo mass bin.

\begin{figure*}
\centering
\resizebox{1.0\textwidth}{!}
{\includegraphics{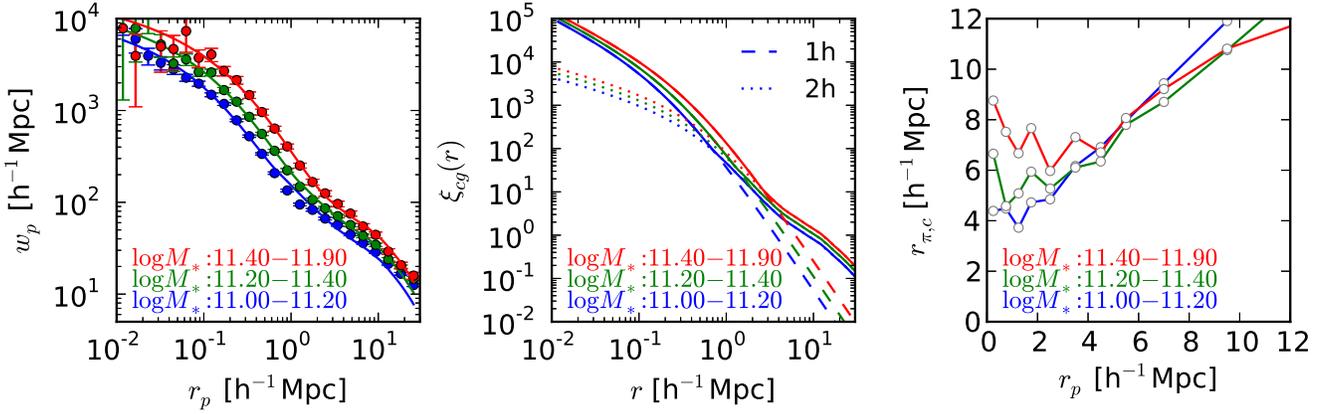}}
\caption{Projected correlations $w_p$~(left), 3D real space
correlations $\xicgr$~(middle), and characteristic U-shaped curves~(right)
for SDSS groups of different BCG stellar mass. {\it Left}: Circles with
error bars show the measurements of $w_p$ from SDSS and solid curves are the
best--fit models.  {\it Middle}: Solid curves show the $\xicgr$ inverted
from best--fit $w_p$ in the left panel. Dashed and dotted curves show the
contributions from the 1-h and 2-h terms in the model. {\it Right}: Same as
the yellow curves shown in Fig.~\ref{fig:xicgsdss}.}
\label{fig:wpinv}
\end{figure*}

\begin{figure*}
\centering
\resizebox{0.9\textwidth}{!}
{\includegraphics{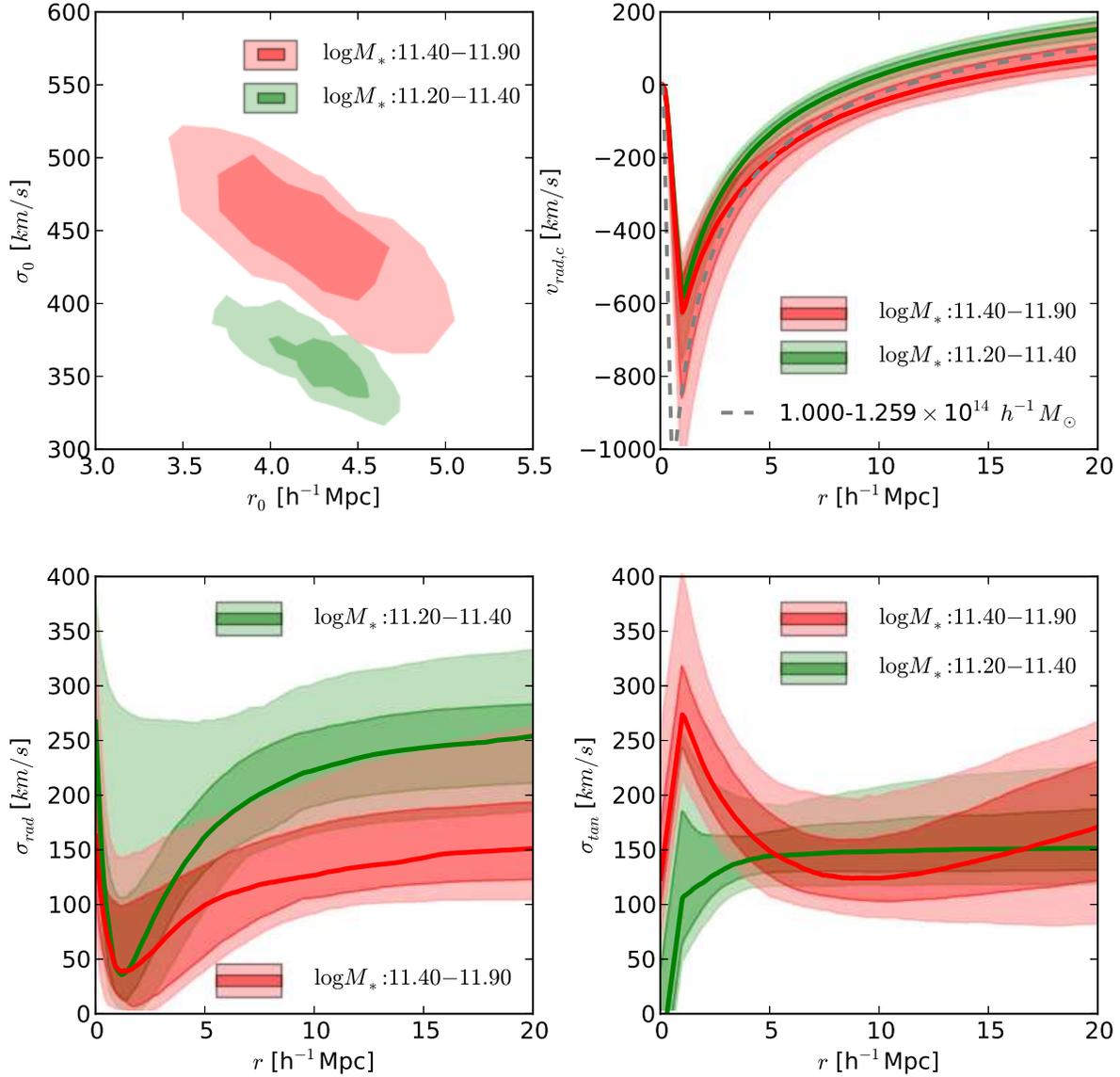}}
\caption{Similar to Fig.~\ref{fig:mcmcmill}, but for the $\xicgs$ measurements
from SDSS groups with BCG $\log\,M_{*}/M_\odot=11.2-11.4$~(green contours) and
$11.4-11.9$~(red contours). The dashed curve in the top right panel shows the
characteristic infall velocity of clusters in mass bin
$1.0\--1.259\times10^{14}\,\hmsol$, measured from the Millennium simulation.}
\label{fig:mcmcsdss}
\end{figure*}

\begin{figure*}
\centering
\resizebox{1.0\textwidth}{!}
{\includegraphics{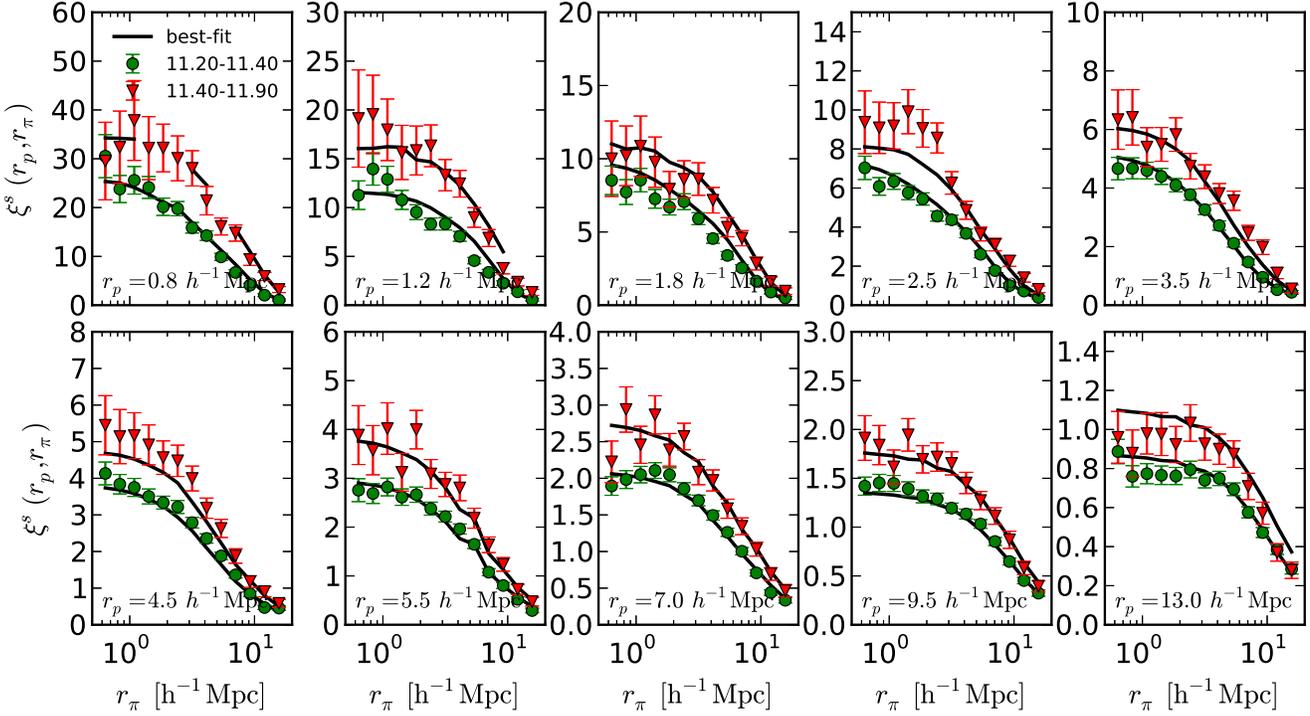}}
\caption{Comparison between the predicted and measured $\xicgs$ at ten different
projected separations for SDSS groups. Green circles and red triangles with
error bars show the measurements for the two BCG stellar mass bins~(marked
in the top left panel), respectively. Solid curves show the predictions from
the best--fit GIK model.}
\label{fig:cutspred2}
\end{figure*}

The first step is to reconstruct $\xicgr$ from the measurements of $w_p$. We
adopt the $\xicgr$ prescription from~\cite{zu2012}. This prescription,
similar to that of~\cite{hayashi2008}, sets the value of $\xicgr$ at each $r$
as the maximum of a Navarro--Frenk--White profile~\citep[][NFW]{navarro1996,
navarro1997} and a biased non--linear matter correlation function. We refer to
these two components as ``1-h'' (for one--halo) and ``2-h'' (for two--halo),
respectively. To ensure a reasonable behaviour of $\xicgr$ on large scales,
we fix the non--linear matter correlation to be that from a flat $\Lambda$CDM
$(\Omega_m=0.23, \sigma_8=0.79)$ universe at $z=0.1$, computed from the
HALOFIT model~\citep{smith2003}. Therefore, the $\xicgr$ model has three
parameters, including cluster bias, scale radius, and normalization of the
NFW profile. For more details about the modeling and calibration of $\xicgr$,
we refer the readers to~\cite{zu2012}. We compute $w_p$ by integrating $\xicgr$
using the same $r_\pi^\mathrm{max}=40\,\hmpc$ and fit it to the measured $w_p$
for each cluster bin. As pointed out by~\cite{croft1999} and~\cite{li2012},
although this reconstruction does not necessarily recover the correct 1-h
and 2-h terms, it provides a reasonable estimate of the underlying $\xicgr$.

Fig.~\ref{fig:wpinv} shows the results of this reconstruction. The left panel
compares the measured $w_p$ to the best--fits using the 3--parameter model,
showing good agreement for the two highest stellar mass bins, with some
discrepancies on large scales for the lowest bin. These discrepancies may be
caused by deficiency of our simplified $w_p$ model, or by contamination from
interlopers in the group catalog. To avoid further complications, we drop the
lowest bin from our following analysis. The middle panel shows the best--fit
$\xicgr$ for each bin, and the corresponding 1-h and 2-h terms. The curves
for the two higher stellar mass bins will be the input $\xicgr$ for our
Bayesian inferences below.  In the right panel, we put together the three
U-shaped curves shown individually in Fig.~\ref{fig:xicgsdss}. They are
qualitatively similar to the curves we see in simulation, and the variation
with BCG stellar mass is as expected --- higher mass bins have stronger
infall and larger dispersion of virial motions, thus smaller $\rpic$ on
large scales but larger $\rpic$ on small scales.

Given the best--fit $\xicgr$ reconstructed from $w_p$, we apply
the same Bayesian inference described in~\S\ref{sec:mcmcmill} to the
measurements of $\xicgs(r_p, r_\pi)$ for each of the two higher stellar
mass bins of SDSS clusters. We vary the same set of parameters listed
in Equation~\ref{eqn:varying}. All the input data points of $\xicgs$
are shown in Fig.~\ref{fig:cutspred2}~(discussed further below) as red
triangles and blue circles with error bars.  Fig.~\ref{fig:mcmcsdss}
presents the constraints on the average GIK for the two bins. The overall
results are remarkably similar to what we see in the simulation~(compare
to Fig.~\ref{fig:mcmcmill}), but with some anomalies that likely arise
from observational uncertainties. In particular, the two inferred values
of $r_0$~(radius at which $\f=1/e$) are considerably larger than expected
for clusters with $M\sim 10^{14}\,\hmsol$~(e.g., $r_0<3\hmpc$ for all the
mass bins in Fig.~\ref{fig:parfit}). This difference is likely caused by
a combination of mis--centering, contamination, and scatter between BCG
stellar mass and cluster mass, all of which blur the $\xicgs$ measurements
on small scales, mimicking a much stronger virial component. However,
even though we do not model these systematic effects, we find trends of
each GIK component with stellar mass that track the trends with cluster
mass seen in the simulation~(Fig.~\ref{fig:parfit}). In particular, the
high stellar mass bin has a higher amplitude of virial dispersion $\s0$,
stronger infall $\vrc$, smaller dispersion in radial velocities $\sr$, and
higher~(comparable) dispersion in tangential velocities $\st$ on small~(large)
scales. The dashed curve in the upper right panel of Fig.~\ref{fig:mcmcsdss}
repeats the black points in Fig.~\ref{fig:parfit}a, marking the infall curve
$\vrc(r)$ measured for $1.0\--1.259 \times 10^{14}\,\hmsol$ halos in the
Millennium simulation. The agreement with our SDSS measurement for the higher
stellar mass bin is very good, indicating that the mass scale of these BCG
$\log\,M_*/M_\odot=11.4\--11.9$ groups is about $10^{14}\,\hmsol$, while
that of the $\log\,M_*/M_\odot=11.2\--11.4$ groups is lower. \citep{li2012}
obtained a similar mass estimate for the $\log\,M_*/M_\odot=11.4\--11.9$ groups
using the internal satellite kinematics~(see their fig. 10).

Fig.~\ref{fig:cutspred2} compares the $\xicgs$ measured from the SDSS to that
predicted from our best--fit models at ten different $r_p$ for the two stellar
mass bins. The model provides a good overall fit to the measurements from
$r_p=0.8\,\hmpc$ to $13\,\hmpc$ in the perpendicular direction, and from
$r_\pi=0.3\,\hmpc$ to $r_\pi=26\,\hmpc$ in the LOS direction for each $r_p$.

\section{Conclusion}
\label{sec:con}

We have developed a methodology for modeling the redshift--space
cluster--galaxy cross--correlation function $\xicgs(r_p, r_\pi)$, calibrating
and testing it with halo and galaxy catalogs from the Millennium simulation
and presenting a first--cut observational application to galaxy groups
in the SDSS redshift survey. The crucial input to this modeling is the
line--of--sight velocity distribution $\fvlos$, which we derive from a
more complete description of the galaxy velocity distribution $\p2d$ that
we refer to as the GIK~(galaxy infall kinematics) model.  Our GIK model
is a 2D mixture of one virialized component~(Gaussian, with zero means
and equal dispersions of radial and tangential velocities) and one infall
component~(bivariate $t$-distribution skewed along the radial velocity
axis). This 2D mixture correctly accounts for the higher moments of the
velocity distributions~(skewness and kurtosis) and the internal correlation
between radial and tangential velocities, providing an excellent fit to
the galaxy kinematics around simulated clusters from the inner $1\,\hmpc$
to beyond $40\,\hmpc$. After convolution with the real--space cluster--galaxy
correlation function, the GIK model accurately reproduces the redshift--space
cluster--galaxy correlation function $\xicgs$ measured in the simulation.

The features of $\xicgs$, which we summarized by the
characteristic U-shaped curve $r_{\pi,c}(r_p)$, are shaped by the complex
interplay among the four distinct elements of the GIK model: the virialized velocity
sphere, the characteristic radial infall velocity, and the radial and tangential
velocity dispersions of the infall component. However, each of these elements
 affects $r_{\pi,c}(r_p)$ differently, and using the Millennium mock data we have
demonstrated that the $\xicgs$ measurement {\it alone} is sufficient to allow 
reconstruction of the underlying GIK around clusters. 
We are especially interested in the characteristic infall curve $\vrc(r)$, as we
expect it to provide a diagnostic of extended cluster mass profiles that is
insensitive to galaxy formation physics that might affect velocity dispersions
within halos. As a proof of concept, we measure $\xicgs$ for SDSS groups and
apply our modeling to infer the GIK for two bins of BCG stellar mass. The four
GIK components show the trends expected if total halo mass correlates with BCG
stellar mass, and the infall curve $\vrc(r)$ for the higher mass bin is in
excellent agreement with the Millennium simulation prediction for
$10^{14}\,\hmsol$ halos.

In principle, the full galaxy pairwise velocity distribution~(as
function of galaxy properties), probed by the redshift--space galaxy
auto--correlation function $\xiggs$, contains more information
than available in $\xicgs$. Current theoretical efforts, both in
configuration space~\citep{tinker2006, tinker2007, reid2011} and in Fourier
space~\citep{seljak2011, okumura2012-1, okumura2012, vlah2012, gil-marin2012,
zhang2012}, are converging to percent--level accuracy of modeling $\xiggs$
on linear scales~(and a few percent on quasi--linear scales), but they are
significantly worse on the non--linear scales where measurements are the most
precise~\citep{scoccimarro2004}. We have shown that galaxy infall onto clusters
is relatively straightforward to model, thanks to the deep potential and high
clustering bias of cluster mass halos. While clusters are rare compared to
galaxies, increasing measurement shot noise relative to $\xiggs$, the high bias
of clusters boosts the signal, so the loss of statistical power may be limited,
and high mass halos are an interesting population to isolate in any case. The
{\it eROSITA} satellite~\citep{merloni2012} and deep optical imaging surveys
from the Dark Energy Survey~\citep{the_dark_energy_survey_collaboration2005}
and LSST~\citep{lsst_science_collaboration2009} should detect $\sim 10^5$
clusters above a $10^{14}\,M_\odot$ threshold, so with an overlapping galaxy
redshift survey the potential measurement precision for $\xiggs$ is very high.

In future work we will investigate the sensitivity of GIK to galaxy formation
physics, including the dependence on large scale spatial bias and velocity
dispersion biases within halos. In observational studies with sufficient
statistics, one can test for systematics by checking that different galaxy
samples~(e.g., blue vs. red) lead to the same cosmological conclusions. The main
observational systematics are scatter between the cluster observable and mass
and mis--centering of clusters~(in both angular and redshift positions). The
large radius of the virial component that we find for our SDSS groups is likely
a consequence of mis--centering effects. Cosmological analyses can incorporate
scatter and mis--centering into the model predictions and marginalize over
uncertainties in their description. However, mis--centering effects must be
calibrated for any given cluster finding algorithm and observational data set
with detailed simulations.

As a probe of dark energy and modified gravity, GIK modeling of galaxy clusters
complements stacked weak lensing analysis in both observational requirements and
information content. Stacked weak lensing relies on overlap between a cluster
sample and a deep imaging survey; forecasts for Stage III and Stage IV dark
energy experiments predict cluster weak lensing constraints that are competitive
with those from supernovae, baryon acoustic oscillations, and cosmic
shear~(see~\citeauthor{weinberg2012} 2012, \S6 and \S8.4). GIK analysis requires
overlap with a large galaxy redshift survey, such as the SDSS survey used here,
the ongoing Baryon Acoustic Oscillation Spectroscopic
Survey~\citep[BOSS;][]{eisenstein2011, dawson2012} and its higher redshift
successor eBOSS, and the deeper surveys planned for future facilities such as
BigBOSS~\citep{schlegel2009}, DESpec~\citep{abdalla2012}, the Subaru Prime Focus
Spectrograph~\citep{ellis2012}, {\it Euclid}~\citep{laureijs2011}, and
{\it WFIRST}~\citep{green2012}. A combinaton of stacked weak lensing and
$\xiggs$ analysis for the same cluster sample would yield tigher dark energy
constraints than either method on its own. Comparison of the two provides an
important consistency test for GR, as many modified gravity models predict a
``slip'' between the gravitational potentials that govern weak lensing and
non--relativistic tracers~\citep[][and references therein]{jain2010}. Compared
to $\Lambda$CDM+GR, modified gravity models predict distinctive signatures in
halo statistics~\citep{chan2009, schmidt2009-1, li2011} and galaxy
redshift--space distortions~\citep{jennings2012}. Most interesting of all for
our purposes, screening effects in modified gravity models lead to distinctive
signatures in halo density profiles~\citep{lombriser2012} and galaxy
phase--space density profiles around clusters~\citep{lam2012}. Redshift--space
cluster--galaxy cross--correlations may be an especially sensitive diagnostic of
such effects, so they could allow stringent tests of these theories, or even
yield smoking gun evidence that deviations from GR on cosmological scales drive
the accelerating expansion of the universe.

\section*{Acknowledgements}

We thank Chris Miller for stimulating discussions about the velocity caustic
method at the initial stage of this project. We also thank Xiaohu Yang for
providing the SDSS group catalog, and thank Cheng Li for discussions on the
stellar mass estimates in SDSS. Y.Z. acknowledges the hospitality of KIPAC
at SLAC \& Stanford University where he enjoyed a fruitful discussion with
participants of the Small Scale Cosmology Workshop. D.H.W. and Y.Z. are
supported by the NSF grant AST-1009505. Y.Z. is also supported by the Ohio
State University through the Distinguished University Fellowship.

The Millennium Simulation databases used in this paper and the web application
providing online access to them were constructed as part of the activities
of the German Astrophysical Virtual Observatory.

Funding for the SDSS and SDSS-II has been provided by the Alfred P. Sloan
Foundation, the National Science Foundation,
the U.S. Department of Energy, the National Aeronautics and Space
Administration, the Japanese Monbukagakusho, the Max Planck Society, and the
Higher Education Funding Council for England, and the Participating
Institutions, which are listed at the SDSS Web Site, http://www.sdss.org/.


\footnotesize{
\bibliographystyle{mn2e}

}

\end{document}